\documentclass[twocolumn, letterpaper, 9.5pt]{article}

\DeclareMathSizes{10}{10.5}{6.5}{6.5}
\usepackage{anysize}
\usepackage{fancyhdr}
 
\marginsize{1.75cm}{1.75cm}{0.5cm}{0.5cm}
 \usepackage[inkscapelatex=false]{svg}
 
\usepackage[square, numbers]{natbib}
\usepackage{easyReview}
\usepackage[utf8]{inputenc} % allow utf-8 input
\usepackage[T1]{fontenc}    % use 8-bit T1 fonts
\usepackage{hyperref}
\usepackage{url}            % simple URL typesetting
\usepackage{booktabs}       % professional-quality tables
\usepackage{amsfonts}       % blackboard math symbols
\usepackage{nicefrac}       % compact symbols for 1/2, etc.
\usepackage{microtype}      % microtypography
\usepackage{amsthm}
\usepackage{caption}
\usepackage{subcaption}
\usepackage{graphics} % for pdf, bitmapped graphics files
\usepackage{epsfig} % for postscript graphics files
\usepackage{times} % assumes new font selection scheme installed
\usepackage{amsmath} % assumes amsmath package installed
\usepackage{amssymb}  % assumes amsmath package installed
\usepackage{makeidx}
\usepackage{float}
\usepackage{booktabs}
\usepackage{mathrsfs}
\usepackage{enumerate}
\usepackage{algorithm}
\usepackage{algpseudocode}

\usepackage{mathrsfs}  
\usepackage{xr} % enable referencing labels in other documents
\usepackage{multicol}
\usepackage{float}
\usepackage[nolist]{acronym}
\usepackage{xcolor}

\colorlet{siaminlinkcolor}{green!50!black}
\colorlet{siamexlinkcolor}{red!50!black}

\usepackage{lipsum} %this package can be removed

\usepackage{cleveref} %Cleveref must be defined after hyperref

\usepackage[acronym]{glossaries} %for use of \gls acronyms

\usepackage{soul}
\usepackage{braket}
\usepackage[compat=0.4]{yquant}
\usetikzlibrary{quantikz2}

\usepackage{scalerel}

\usepackage{xcolor}

\begin{document}

\title{
	\vspace{-1.5cm}\rule{\linewidth}{4pt}\vspace{0.3cm} \Large \textbf{
    % Hardware-Demonstrated Quantum Echo State Networks for Predicting Chaotic Systems
    Quantum Observers: A NISQ Hardware Demonstration of Chaotic State Prediction Using Quantum Echo-state Networks
    %Quantum Echo-State Networks for Predicting Chaotic Dynamics on NISQ Hardware
    % \cred{A NISQ Hardware Demonstration} of Chaotic System \\ Prediction with Quantum Echo-state Networks
    % Predicting Chaotic Systems with Quantum Echo-state Networks
	}\\ \rule{\linewidth}{1.5pt}}
	\author{Erik L. Connerty\thanks{Corresponding author. Email: \href{erikc@cec.sc.edu}{erikc@cec.sc.edu}}, 
     Ethan N. Evans, Gerasimos Angelatos, and Vignesh Narayanan\\ 
    \vspace{-.0cm}
	\small{University of South Carolina - Columbia} \vspace{-0.0cm} \\ \small{Naval Surface Warfare Center, Panama City Division} \vspace{-0.0cm} \\ \small{RTX BBN, Cambridge MA} \vspace{-0.0cm}
	}
        \date{\vspace{-.5cm}}
	\maketitle

\begin{abstract}
Recent advances in artificial intelligence have highlighted the remarkable capabilities of neural network (NN)-powered systems on classical computers. However, these systems face significant computational challenges that limit scalability and efficiency. Quantum computers hold the potential to overcome these limitations and increase processing power beyond classical systems. Despite this, integrating quantum computing with NNs remains largely unrealized due to challenges posed by noise, decoherence, and high error rates in current quantum hardware. Here, we propose a novel \emph{quantum echo-state network} (QESN) design and implementation algorithm that can operate within the presence of noise on current IBM hardware. We apply classical control-theoretic response analysis to characterize the QESN, emphasizing its rich nonlinear dynamics and memory, as well as its ability to be fine-tuned with sparsity and re-uploading blocks. We validate our approach through a comprehensive demonstration of QESNs functioning as quantum observers, applied in both high-fidelity simulations and hardware experiments utilizing data from a prototypical chaotic Lorenz system. Our results show that the QESN can predict long time-series with persistent memory, running over $100$ times longer than the median $\mathcal{T}_1$ and $\mathcal{T}_2$ of the \texttt{ibm\_marrakesh} QPU, achieving state-of-the-art time-series performance on superconducting hardware.

\end{abstract}

\maketitle

%=============== Section: Intro ==================
\section{Introduction}\label{sec: Introduction}
%=============== Section: Intro ==================
% Quantum computers are poised to offer a unique framework for the adaptation of neural networks and all of their offshoots into quantum versions of their classical representations. Due to the inherent difficulties of implementing a large number of quantum gates, noise, and the short coherence times of current noisy intermediate-scale quantum (NISQ) hardware \cite{gill2022quantum}, we sought to develop a foolproof method for running long time-series prediction on NISQ hardware and beyond. Our work aims to advance some of these challenges by presenting a scalable recurrent neural network (RNN) architecture for both present NISQ and future fault-tolerant quantum computers by leveraging design-principles from the classical echo-state networks (ESNs).
Quantum computers are poised to offer a unique and powerful computational framework, capable of transforming the current machine learning paradigm into quantum counterparts and unlocking new levels of efficiency and power that could redefine computational models. However, these capabilities are not yet fully realized due to significant implementation challenges, including short coherence times, noise, measurement backaction, and the inherent difficulties of scaling up the number of quantum gates—limitations that hinder the effectiveness of current noisy intermediate-scale quantum (NISQ) hardware, particularly in the development of quantum neural networks \cite{gill2022quantum}. Despite the limitations of NISQ devices in comparison to fault-tolerant quantum computers, which are expected to offer far greater reliability and scalability, the NISQ hardware still provide a valuable testbed for developing and testing quantum neural networks and learning algorithms. 

Overcoming the challenges due to decoherence, noise, and high-error rates when implementing a large-number of quantum gates have been the primary focus of efforts towards quantum machine learning on NISQ. In this context, quantum neural networks, particularly, quantum reservoir networks (QRNs) have garnered significant attention. Reservoir networks, in classical computing, are a class of recurrent neural networks employed to model relations between sequential data. However, developing quantum versions of classical machine-learning algorithms is very rarely accomplished by converting instructions over one-to-one \cite{evans2024learningsasquatchnovelvariational, cnn_Cong_2019, gnn_verdon2019quantumgraphneuralnetworks, qrl_Daoyi_Dong_2008, qdiff_cacioppo2023quantumdiffusionmodels}. For example, in the context of reservoir networks, one must reproduce notions such as fading memory, asymptotic stability (called the ``echo-state'' property), and \emph{nonlinearity} in QRNs. These present important challenges to overcome due to the inherent difficulty of quantum coherence times \cite{gill2022quantum} and state measurement collapse, as well as the inherent unitary evolution of closed quantum systems. % which may naively seem to prevent the implementation of nonlinear operations.
    
% ESNs \cite{jaeger2001}, a class of RNNS, offer a unique way of handling time-series data. Specifically, without the need for backpropagation techniques or vast amounts of labeled data, ESNs can perform challenging tasks with very little hyperparameter tuning. Moreover, sparsely connected hidden layers within ESNs \cite{jaeger2001,miao2022interpretable, sparse_YANG202295,sparse_gallicchio2020sparsityreservoircomputingneural} present a desirable feature that can reduce the gate complexity of the quantum circuit (QC) and may prove to be a promising feature to implement on current-day NISQ computers. However, developing a quantum version of a classical machine-learning algorithm very rarely is accomplished by converting instructions over one-to-one \cite{evans2024learningsasquatchnovelvariational, cnn_Cong_2019, gnn_verdon2019quantumgraphneuralnetworks, qrl_Daoyi_Dong_2008, qdiff_cacioppo2023quantumdiffusionmodels}. For example, in the context of ESNs, one must reproduce notions such as fading memory, asymptotic stability (called the``echo-state'' property), and \emph{nonlinearity} in quantum reservoir networks (QRNs). These present important challenges to overcome due to the inherent difficulty of quantum coherence times \cite{gill2022quantum} and state measurement collapse, as well as the inherent unitary evolution of closed quantum systems, which may naively seem to prevent the implementation of nonlinear operations. 
%=========== Figure 1 ===========
\begin{figure*}[t]
    \centering
    \includegraphics[width=\linewidth,keepaspectratio]{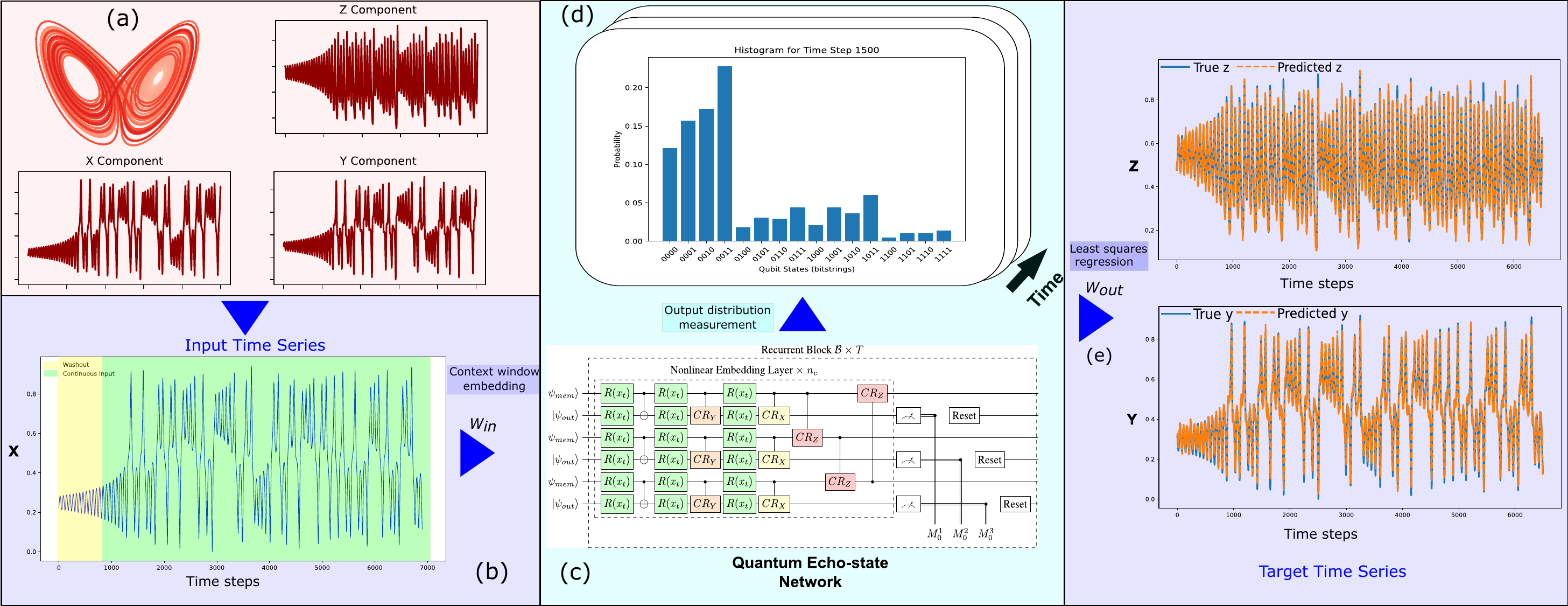}
    \caption{Proposed QESN observer framework.(a) Phase and state trajectories of the Lorenz system operating in a chaotic regime. (b) X-coordinate data is input as a sliding context window over time through a fully connected classical layer $W_{in}$, where part of the time window is marked as washout during which the output is not processed and (c) then passed through a dynamic quantum reservoir that is repeatedly sampled; (d) the output probability distribution sampled over the $2^\frac{n_q}{2}$-dimensional Hilbert space at each time step. With the collected probability distributions, a classical least-squares optimization process fits the reservoir states to the target signals (e) corresponding to the $Y$ and $Z$ coordinates of the Lorenz system.}
    \label{fig:QRC_Pipeline}
\end{figure*}
%=========== Figure 1 ===========
% \begin{figure}
%     \centering
%     \includegraphics[width=\linewidth]{figures/lorenz-multiplot.pdf}{}
%     \caption{The Lorenz `63 System.}
%     \label{fig:lorenz}
% \end{figure}
% %=========== Figure 2 ===========
%=========== Figure 2 ===========

%-------------- Related works ------------
% \textbf{Related Works:}
%-------------- Related works ------------
Previous research has examined the creation of functional quantum recurrent neural networks (QRNNs) as well as QRCs. One such approach, detailed in \cite{LI2023148}, developed a QRNN trained through the backpropogation algorithm. Similar to the classical RNNs, QRNNs inherit high training costs, suffer from vanishing gradients, and require large number of circuit evaluations. In addition, several studies (e.g., \cite{kornjača2024largescalequantumreservoirlearning, Govia_2022, zhu2024practicalscalablequantumreservoir, dudas2023, Mujal2023, PhysRevResearch.4.033007}) have explored the use of analog quantum computers for time-series prediction within open quantum systems. However, these approaches are limited to hardware-specific, specialized non-universal devices. In contrast, focusing on universal quantum devices, recent efforts \cite{Hu2024, yasuda2023quantumreservoircomputingrepeated, chen2020temporal} have introduced frameworks and techniques that allow quantum computers to operate effectively over extended durations without the need for intermediate halts or system resets. % On the other hand, focusing on universal devices, efforts in \cite{Hu2024, yasuda2023quantumreservoircomputingrepeated, chen2020temporal} present algorithms that enable quantum computers to operate seemingly indefinitely without intermediate halts or system resets. 
Specifically, controlled input-dependent rotations on measured qubits, along with deterministic resets, were incorporated to implement a guaranteed weak-entanglement scheme. The Volterra kernel analysis technique was employed to demonstrate that this \textit{measure and reset paradigm} enhances circuit stability and facilitates a controlled destruction of information at each time step, thereby enabling a time-invariant fading memory through the implementation of a controlled weak-entanglement scheme \cite{Hu2024}.
%In particular, controlled input-dependent rotations on measured qubits as well as deterministic resets work were incorporated to create a guaranteed weak-entanglement scheme that stabilizes the circuit, as well as destroys information at each time-step, enabling a time-invariant fading memory. %\cred{add current status on hardware experiments.. something like 'despite theoretical evidence.... existing or state-of-the-art implementations...'}. \cred{Add methods that are used to analyze nonlinearity of memory in existing networks... including the type of analysis reported on the effect of gates, memory, repeating reuploading, input context/encoding..}
These methods, while providing important theoretical foundations for developing QRCs, often lack thorough hardware validation due to noted technical issues such as “buffer overflows” \cite{Hu2024}, lack of persistent memory \cite{kornjača2024largescalequantumreservoirlearning}, or measuring paradigms that require circuit re-initialization \cite{Mujal2023}, making them more computationally expensive and less practical. %Additionally, these approaches generally do not support tunable nonlinearity, a key feature that is essential for modeling chaotic systems, and a comprehensive analysis of QRC output features and their properties remains absent in the literature.

To address these gaps, and inspired by the emerging {measure and reset} paradigm, we propose a scalable quantum echo state network (QESN) circuit. The proposed QESN is used to develop a \textit{quantum observer},  a virtual sensor that reconstructs latent state variables from measured outputs of a dynamical system. These observers can be implemented on any digital quantum computer implementing a universal gate set, and do not require any sort of stopping or re-initialization of the circuit, allowing continuous evolution of the quantum state over long time horizons. To achieve this, our design introduces a context window via angle embeddings and \textit{sparsity} as a key feature, implemented via randomly initialized input layer weights and a sparsely connected reservoir layer—paired with a robust and tunable entangling mechanism. This architecture enables a definitive demonstration of long time-series prediction on digital quantum computers. To support circuit design, we also introduce a detailed response analysis framework inspired by classical control theory, which we use to analyze the effect of sparsity on the nonlinearity and memory capacity of the QESN. To validate the capabilities of the QESN, we deploy it as a dynamic quantum observer tasked with reconstructing the latent state trajectories of a Lorenz system exhibiting chaotic dynamics. %We report experimental results where the designed quantum observer achieve state-of-the-art performance on current NISQ hardware, demonstrating the powerful potential of today’s quantum devices. 
We report experimental results in which the designed quantum observer achieves state-of-the-art performance on contemporary NISQ hardware, underscoring the remarkable potential of today’s quantum devices. Specifically, our results demonstrate accurate time-series prediction across durations $100$ times longer than the median $\mathcal{T}_1$ and $\mathcal{T}_2$ coherence times of the qubits—surpassing all prior demonstrations to date.

\section{Methods}\label{sec:methods}
%================= sec: Methods ===================
We propose a near term NISQ ready algorithm that can leverage the large Hilbert space afforded by a set of qubits to produce a rich feature space that is suitable for predicting complex dynamics. Our QESN observer features a fully connected input layer utilizing a ``context window'' for data input, a novel entanglement scheme that uses controlled rotations and the ``data reuploading'' \cite{P_rez_Salinas_2020} technique, and randomly generated weights throughout the circuit.  Figure \ref{fig:QRC_Pipeline} details the entire proposed quantum observer pipeline, % from start to finish, 
showing how classical data is input into the QESN, evolved through a quantum channel, and then sampled to generate features for a classical optimization problem.

\paragraph{Angle Embedding:}

Qubits are easily represented geometrically using a spherical object known as the Bloch sphere \cite{blochfeyn/10.1063/1.1722572, blochPhysRevA.5.1094}, with pure quantum states existing on the surface of this sphere. Naturally, to embed data onto this surface, angle embeddings, which use \textit{angled rotations}, are a natural choice. In the context of the QESN, angle embeddings are used to encode classical data onto the quantum mechanical Hilbert space generated by quantum computers. In contrast to an ``amplitude embedding'', a method that encodes data into complex probability amplitudes directly, angle embeddings can be used to continuously evolve quantum states, an attribute which allows for online reservoir computing with persistent memory. Let $R_x$ and $R_z$ be arbitrary single-qubit rotation gates along their respective Pauli basis, and let the three variables $\alpha,\beta,\gamma$ be the three Euler angles needed for an arbitrary rotation around the Bloch sphere. Then, the arbitrary rotation gate $R$ can be implemented as a sequence of rotations $(\alpha, \beta, \gamma)$, given by 
\begin{equation}
R(\alpha, \beta, \gamma) = R_z(\gamma) R_x(\beta) R_z(\alpha).
\end{equation}
This arbitrary rotation gate provides a mapping from classical data into a single-qubit rotation. While a large set of valid Euler angles exist, the sequence $ZXZ$ was chosen due to the ability to have virtualized $Z$ rotations with zero error \cite{PhysRevA.96.022330} on IBM hardware.

\paragraph{Context Window:}  {We implement a method for embedding a context window of data using a fully connected input layer. This embedding technique maps arbitrary sized inputs into three Euler angles for each qubit, and allows us to decouple the size of the reservoir from the dimension of the input}. Let the input vector be $x_t \in \mathbb{R}^{d}$,  $t \in T = \{1, \dots, N\}$. We encode the classical data onto $n_q$ total qubits, using a context window $X_c = \{x_{t-c}, \dots, x_t\}$ of length $c$. Let the input weight tensor and the bias be denoted by $W \in \mathbb{R}^{cd \times n_q \times 3}$ and $b \in \mathbb{R}^{n_q}$, respectively. The output rotation angles be $\Theta \in \mathbb{R}^{n_q \times 3}$ for each qubit $i$ along the three axes indexed by $j$, is computed as the dot product between the input vector $x$ and the corresponding weight slice $W_{\cdot, i,j}$, and then add the bias term $b_i$.
\begin{align}\label{eq:input-embedding}
\Theta_{ij} = \sum_{t=1}^{cd} x_t \, W_{t,i,j} + b_i,
\end{align}
for $j=1,2,3$, and $i=1,\ldots, n_q$.
% for each qubit  $i$. 

%=========== Figure 1 ===========
\begin{figure*}
    \centering
    \resizebox{\textwidth}{!}{% \usetikzlibrary{decorations.pathreplacing, positioning, decorations.pathmorphing}
% \usetikzlibrary{fit, shapes.geometric}

\begin{tikzpicture}[shorten >=1pt, node distance=2cm and 3cm]
    \tikzstyle{unit}=[draw,shape=circle,minimum size=1.1cm,fill=white]
    % \tikzstyle{output}=[draw,shape=circle,minimum size=1.1cm,fill=white]
    % % Input neurons positioned more to the left
    % \node[unit](x00) at (-5.4,.4){};
    % \node[unit](x01) at (-5.4,-1.6){};
    % \node[unit](x0d) at (-5.4,-3.9){};
    % \node[unit](x10) at (-5.2,.2){};
    % \node[unit](x11) at (-5.2,-1.8){};
    % \node[unit](x1d) at (-5.2,-4.1){};
    % \node[unit](x20) at (-5,0){$x_{0,t}$};
    % \node[unit](x21) at (-5,-2){$x_{1,t}$};
    % \node at (-5.3,-2.8){\vdots};
    % \node[unit](x2d) at (-5,-4.3){$x_{D,t}$};

    % Quantum circuit as the "hidden layer"
    \begin{scope}[shift={(-2.65,0)}]
      \begin{yquant*}%[every post measurement control=direct]%[every multi label/.style={every node/.style={anchor=east, midway}}]%[operators/every h/.append style={fill=red!20}]
    
    % \yquantset{every multi label/.style={every node/.style={anchor=east, midway}}}
    
    % [name=past2] qubit {} t0[2];
    % [name=past1] qubit {} t1[2];
    % [name=past] qubit {} t2[2];

        [name=past2a] qubit {$\ket{\psi_{mem}}$} t0;
        [name=past2b] qubit {$\ket{\psi_{out}}$} t0b;

        [name=past1a] qubit {$\ket{\psi_{mem}}$} t1;
        [name=past1b] qubit {$\ket{\psi_{out}}$} t1b;

        [name=pasta] qubit {$\ket{\psi_{mem}}$} t2;
        [name=pastb] qubit {$\ket{\psi_{out}}$} t2b;
    % cbit {$c$} c[1];
    nobit out;
    
    hspace {1mm} -;
    
    % \node[anchor=east] at($(past2-0.west)!.5!(past2-1.west)$) {$\ket\psi$};
    
    [this subcircuit box style={dashed, "Recurrent Block $\mathcal{B}$ $\times \; T$"}]
    subcircuit {
        [inout] 
        qubit {}t0;
        qubit {}t0b;
        qubit {}t1;
        qubit {}t1b;
        qubit {}t2;
        qubit {}t2b;
        % cbit {} c[1];
        nobit out;

    % qubit {$\psi_{t-2}$} t0[2];
    % qubit {$\psi_{t-1}$} t1[2];
    % qubit {$\psi_{t}$} t2[2];
    % cbit {$c$} c[1];
    % nobit out;
    
    [this subcircuit box style={dashed, "Nonlinear Embedding Layer $\times \; n_c$"}]
    subcircuit {
        [inout] 
        qubit {}t0;
        qubit {}t0b;
        qubit {}t1;
        qubit {}t1b;
        qubit {}t2;
        qubit {}t2b;
    
    [fill=green!20] box {$R(x_t)$} t0; 
    [fill=green!20] box {$R(x_t)$} t0b;
    cnot t0b | t0;
    [fill=green!20] box {$R(x_t)$} t0;
    [fill=green!20] box {$R(x_t)$} t0b;
    [fill=orange!20] box {$CR_Y$} t0b | t0;
    [fill=green!20] box {$R(x_t)$} t0;
    [fill=green!20] box {$R(x_t)$} t0b;
    [fill=yellow!20] box {$CR_X$} t0b | t0;
    
    [fill=green!20] box {$R(x_t)$} t1;
    [fill=green!20] box {$R(x_t)$} t1b;
    cnot t1b | t1;
    [fill=green!20] box {$R(x_t)$} t1;
    [fill=green!20] box {$R(x_t)$} t1b;
    [fill=orange!20] box {$CR_Y$} t1b | t1;
    [fill=green!20] box {$R(x_t)$} t1;
    [fill=green!20] box {$R(x_t)$} t1b;
    [fill=yellow!20] box {$CR_X$} t1b | t1;
    
    [fill=green!20] box {$R(x_t)$} t2;
    [fill=green!20] box {$R(x_t)$} t2b;
    cnot t2b | t2;
    [fill=green!20] box {$R(x_t)$} t2;
    [fill=green!20] box {$R(x_t)$} t2b;
    [fill=orange!20] box {$CR_Y$} t2b | t2;
    [fill=green!20] box {$R(x_t)$} t2;
    [fill=green!20] box {$R(x_t)$} t2b;
    [fill=yellow!20] box {$CR_X$} t2b | t2;
    
    [fill=red!20] box {$CR_Z$} t1 | t0;
    [fill=red!20] box {$CR_Z$} t2 | t1;
    [fill=red!20] box {$CR_Z$} t0 | t2;
    % box {$CR_Z$} t0[0] | t22[0];
    
    } (t0,t0b,t1,t1b,t2,t2b);
    
    measure t0b;
    measure t1b; 
    measure t2b;
    
    % text {$q_0^1, q_0^2, q_0^3$} out | t0[1],t1[1],t2[1];
    text {$M^1_0$} out | t0b;
    text {$M^2_0$} out | t1b;
    text {$M^3_0$} out | t2b;
    
    discard t0b,t1b,t2b;
    % init {Reset} t0b;
    % init {Reset} t1b;
    % init {Reset} t2b;
    box {Reset} t0b;
    box {Reset} t1b;
    box {Reset} t2b;
    % discard t0b,t1b,t2b;
    % init {$\ket{0}$} t0b;
    % init {$\ket{0}$} t1b;
    % init {$\ket{0}$} t2b;
    settype {qubit} t0b;
    settype {qubit} t1b;
    settype {qubit} t2b;
    
    } (t0,t0b,t1,t1b,t2,t2b,out);
    
    %hspace {3.5mm} -;
    % [shape=yquant-rectangle, direct control] measure {$\calR$} (t0,t1,t2);
    % text {$\{q_1^1, q_1^2, q_1^3\}$} out | t0,t1,t2;
    % settype {qubit} t0;
    % settype {qubit} t1;
    % settype {qubit} t2;
    % discard t0[1],t1[1],t2[1];
    
    % hspace {-3mm} -;
    % text {$\dots$} t0,t1,t2;
    % hspace {-3mm} -;
    % [shape=yquant-rectangle, direct control] measure {$\calR$} (t0,t1,t2);
    % text {$\{q_s^1, q_s^2, q_s^3\}$} out | t0,t1,t2;
    % settype {qubit} t0;
    % settype {qubit} t1;
    % settype {qubit} t2;
    
    % hspace {-3mm} -;
    % text {$\dots$} t0,t1,t2;
    % hspace {-3mm} -;
    % [shape=yquant-rectangle, direct control] measure {$\calR$} (t0,t1,t2);
    % text {$\{q_T^1, q_T^2, q_T^3\}$} out | t0,t1,t2;
    % settype {qubit} t0;
    % settype {qubit} t1;
    % settype {qubit} t2;
    % hspace {-4mm} -;
    
    % hspace {1mm} -;
    
    % box {$\text{Recurrent Block}$} (t0,t1,t2);
    % settype {cbit} t0[1];
    % settype {cbit} t1[1];
    % settype {cbit} t2[1];
    % text {$q^1_0$} out | t0[1];
    % text {$q^2_0$} out | t1[1];
    % text {$q^3_0$} out | t2[1];
    % discard t0[1],t1[1],t2[1];
    
    \end{yquant*}
    \end{scope}

\end{tikzpicture}}
    \caption{The QESN Circuit. Two sub-circuits compose the entire block. The interior block is called the ``Nonlinear Embedding Layer'' and encodes the time-series data at time point $\tau$ into the circuit through the pictured unitary. Randomized weights are generated for each run, making the circuit's generation of weights and sparsity dynamic, similar to classical ESNs. The gates denoted as $\mathcal{R}$ are unitary operations that perform arbitrary rotation based off of the input time series data. The inner loop is repeated $n_c$ times and serves as a tunable hyperparameter for increasing input-to-output nonlinearity. The outer block labeled ``Recurrent Block'' serves as the main loop for the QESN algorithm, running once for every point in the input data. A measurement is taken at the end of this block on a subset of the qubits, and then those same qubits are reset to $\ket{0}$.}
    \label{fig:QRC_Circuit}
\end{figure*}
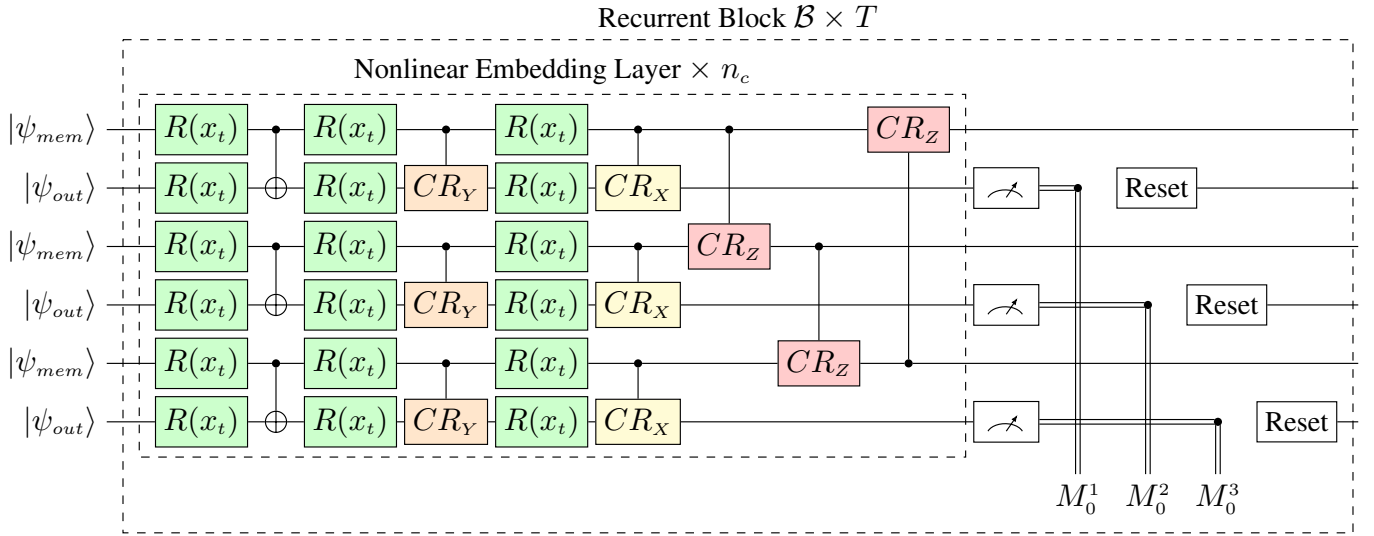
%=========== Figure 1 ===========

\paragraph{QESN Circuit:} For the proposed quantum observer, we design a QESN circuit, which is depicted in Figure \ref{fig:QRC_Circuit}. In the designed circuit, the \emph{memory} and \emph{readout} registers, denoted by $|\psi_{mem}\rangle$ and $|\psi_{out}\rangle$, respectively, are chosen in equal numbers and placed in alternating channels of the circuit. Single qubit rotations interspersed with two-qubit entangling gates are included to increase nonlinearity of the mappings from input to output, as well as to create combinations of previous inputs with current inputs (i.e., memory) \cite{chu2022qmlperrortolerantnonlinearquantum, Govia_2022} and allow the necessary transfer of information from the memory register to the readout register. 
 
Sparsity, an attribute that requires the random deletion of a subset of two-qubit entangling gates in the circuit, is used to decrease the number of gate errors, as gates that are deleted are removed from the circuit altogether. %Additionally, sparsity can reduce the depth of the circuit by decreasing SWAP operations, speeding up transpilation for IBM NISQ hardware. Thus we see sparsity as a critical component in our QESN design, especially for NISQ hardware. 
Additionally, sparsity contributes to reduced circuit depth by minimizing the number of SWAP operations, thereby accelerating transpilation on IBM NISQ hardware. Thus, we utilize sparsity as a critical design component in our QESN architecture, particularly within the context of NISQ-era quantum devices. We also introduce a tunable ``data re-uploading block'' that allows tunable amounts of nonlinearity within the circuit, inspired by \cite{P_rez_Salinas_2020,chu2022qmlperrortolerantnonlinearquantum}. At each timestep, a probability distribution is sampled from the readout qubits of the QESN to create features for a classical least-squares regression with elastic net regularization \cite{elastic_net} to avoid overfitting. By selectively measuring only the readout qubits and then deterministically resetting them to an unbiased state, we are able to run the circuit for an arbitrarily long time while retaining a fading memory \cite{Hu2024,yasuda2023quantumreservoircomputingrepeated,chen2020temporal}. The detailed algorithm for implementing the proposed QESN-based observer in given in Supplementary Note \ref{SN:algorithm}. %Fading memory is an important feature of classical reservoir networks, and has been analyzed in detail using Volterra kernel analysis in \cite{Hu2024}. 
%============ Figure 4 ==============
   \begin{figure*}[h!]
        \centering
        \includegraphics[width=\linewidth, keepaspectratio]{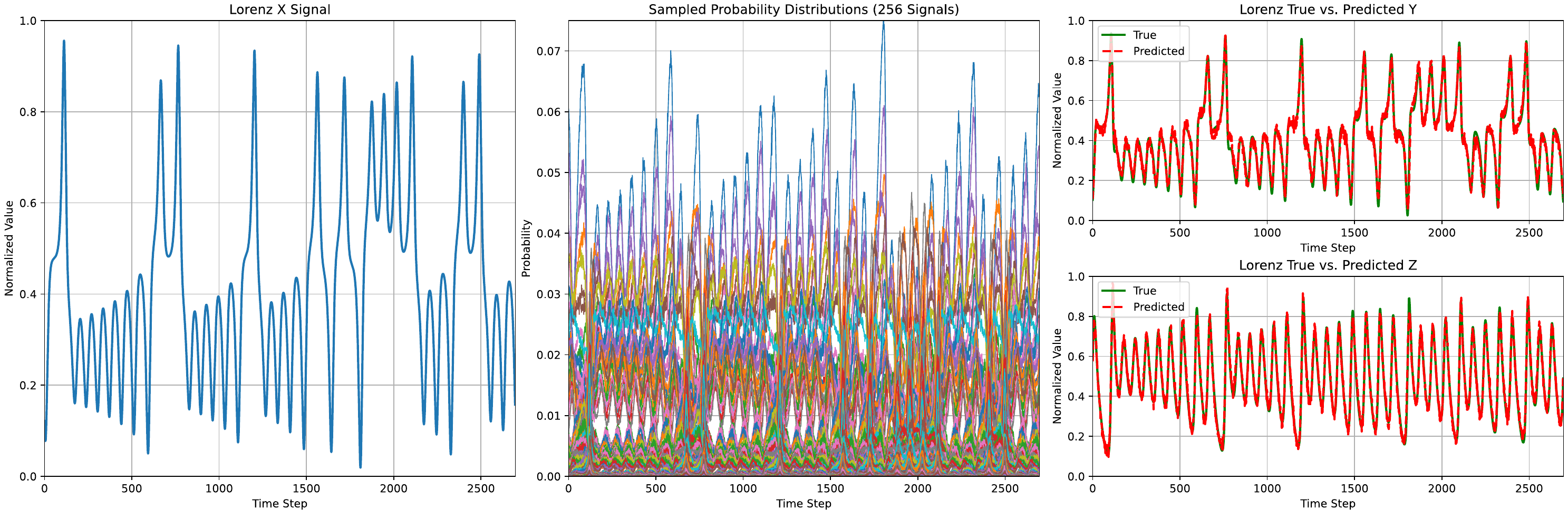}
        \caption{The full predictive results from the Aer simulator on the Lorenz $y(t+1)$ and $z(t+1)$ components. The simulation was done with the QESN circuit with $n_q=16$ qubits. From left-to-right the input signal, sampled reservoir features, and predictions are depicted, respectively. Input data is generated from the chaotic Lorenz system and only includes the $x(t)$ component. The probability distributions of output states ${(2^8)}$ of the QESN are sampled by taking measurements from the QESN readout circuit $60,000$ times at each sampling instant, over the simulation time. At the end, the sampled probability distributions are mapped to the target signals using elastic net regression and the estimated $y(t)$ and $z(t)$ trajectories of the Lorenz system are plotted in the right panel.}
        \label{fig:qrnn_test}
    \end{figure*}
%============ Figure 4 ==============
\paragraph{Response Analysis and Sparsity Tuning:}
In classical control theory, time-domain response analysis serves as a fundamental approach for characterizing the dynamic behavior of systems subjected to specific input signals, such as step, sinusoidal, or ramp functions. This method provides critical insights into system properties, including stability, transient behavior, and steady-state accuracy, often quantified by metrics like rise time, settling time, overshoot, and steady-state error \cite{dorf2005modern}. We extend these classical principles to the quantum domain by applying time-domain analysis to the proposed QESN to examine its dynamic characteristics. By analyzing the temporal evolution of quantum states in response to structured input sequences (step, ramp, and sinusoidal), we characterize the system’s dynamical behavior, revealing rich nonlinear responses as well as its coherence properties. 

These insights inform the memory capacity and the degree of nonlinearity in the input-to-output mapping of the QESN. Moreover, this approach facilitates a systematic evaluation of the QESN's response characteristics, shedding light on its robustness and reliability—particularly as sparsity is introduced into the circuit's gate weights, and as differing number of ``data re-uploading blocks'' are used. For instance, using this method, our experiments show that we can eliminate approximately $50\%$ of the entanglement gate weights without compromising the desired properties of the QESN-based observers, providing a useful empirical approach for sparsity selection. We also find the optimal number of repeatable blocks using this response analysis, setting the number of repeats to be $3$ after a guided examination of different parameters. This response analysis, in turn, guides the selection of the sparsity index for the weight matrices associated with the entanglement gates, as well as the number of repeatable ``data re-uploading blocks'', optimizing gate operations and reducing overall circuit depth. A comprehensive account of the analysis methodology and results is provided in Supplementary Note \ref{SN:response-analysis}.
%In classical control theory, time-domain response analysis is a foundational method for characterizing the dynamic behavior of systems subjected to specific input signals, such as step, impulse, sinusoidal, or ramp functions. This analysis yields critical insights into system properties, including stability, transient behavior, and steady-state accuracy, typically quantified using metrics such as rise time, settling time, overshoot, and steady-state error \cite{dorf2005modern}. In this work, we extend the principles of classical time-domain response analysis to the quantum domain by applying them to the proposed QESN to analyze its dynamic characteristics. By examining the temporal evolution of quantum states in response to structured input sequences (step, ramp, and sinusoidal), we characterize the circuit’s dynamical behavior, revealing rich nonlinear responses, alongside its coherence properties, which inform the memory capacity, and gate-level performance. This approach enables a systematic evaluation of the QESN’s response characteristics, offering insights into its robustness and reliability—particularly as sparsity is introduced into the circuit's gate weights. This, in turn, facilitates an informed selection of the sparsity index for the weight matrices associated with the entanglement gates that minimize gate operations and reduce overall circuit depth. A comprehensive account of the analysis methodology and corresponding results is presented in Supplementary Note \ref{SN:response-analysis}.

%================== sec: results ===================
\section{Simulation Results}\label{sec:results}
%================== sec: results ===================
% \subsubsection{Simulation}

We conducted our simulation experiments using the Aer simulator. Results were gathered with mid-circuit measurements and recovered in two different ways: (a) as expectation values, which provide the mean value of each qubit's Pauli-Z outcome, and (b) as the full probability distribution over the computational basis, yielding $2^{\frac{n_q}{2}}$ features, where $\frac{n_q}{2}$ represents the number of readout qubits. These mid-circuit measurements are accumulated over $60,000$ shots for a single circuit run, ensuring statistical robustness and accuracy of the measured outputs. Sampling the circuit over a large number of shots is extremely important as reproducing the {$2^{\frac{n_q}{2}}$} output probability distribution accurately will generally improve predictive performance. The QESN's performance is then evaluated on its ability to accurately predict the $\big(y(t+1), z(t+1)\big)$ components of the Lorenz system \cite{DeterministicNonperiodicFlow} given only the $x(t)$ component, with special emphasis on the test predictions. This task typically requires a fading memory and nonlinear activation function in classical reservoir networks \cite{miao2022interpretable}.

    %    \begin{figure}[h]
    %     \centering
    %     \includegraphics[width=\linewidth, keepaspectratio,trim={0cm 0 0 1.4cm},clip]{Quantum Reservoir Computing for Time Series Prediction/figures/QESN_pred_vs_true.pdf}
    %     \caption{\footnotesize Test prediction from X to Y component in the Aer simulator with 16 qubits.}
    %     \label{fig:qrnn_test_y}
    % \end{figure}

%============ Table 1 ==============
    \begin{table}[h!]
        \centering
        \small 
        \begin{tabular}{c|c c c}
             Qubits & Expectation & Probability & Distribution\\
              & Value & Distribution & w. Noise\\
             \hline
             \textbf{4 Qubits} &  & \\
             Train RMSE & \textbf{.1124} & .1177 & .1468\\
             Test RMSE & \textbf{.1112} & .1185 & .2016\\
             \hline
             \textbf{6 Qubits} & & \\
             Train RMSE & .0986 & \textbf{.0616} & .1193\\
             Test RMSE & .0963 & \textbf{.064} & .1315\\
             \hline
             \textbf{8 Qubits} &  & \\
             Training RMSE & .0822 & \textbf{.0429} & .1110\\
             Test RMSE & .0798 & \textbf{.0463} & .1285\\
             \hline
             \textbf{10 Qubits} &  & \\
             Train RMSE & .0688 & \textbf{.0425} & .1258\\
             Test RMSE & .0699 & \textbf{.0422} & .1298\\
             \hline
             \textbf{12 Qubits} &  & \\
             Train RMSE & .0631 & \textbf{.0378} & .0986\\
             Test RMSE & .0635 & \textbf{.0377} & .1282\\
             \hline
             \textbf{14 Qubits} &  & \\
             Train RMSE & .0476 & \textbf{.024}  & .0754\\
             Test RMSE & .046 & \textbf{.0249} & .0988\\
             \hline
             \textbf{16 Qubits} &  & \\
             Train RMSE & .0488 & \textbf{.0225} & .0573\\
             Test RMSE & .0493 & \textbf{.0237}  & .0895\\
        \end{tabular}
        \caption{Simulated training and test error using various different feature recovery methods and noise configurations measured in RMSE (Root mean squared error). An \texttt{ibm\_fez} noise model was used to gather the noisy results. The best run from each category was used, and the elastic net regularization parameters were tuned for each bin to get lower test loss.}
        \label{tab:comparison_table}
    \end{table}
%============ Table 1 ==============
The QESN circuit was first tested on the Lorenz system %in simulation using the Aer simulator
with varying numbers of qubits from $4$ to $16$. An NVIDIA DGX-A100 was used to perform circuit simulations and collect results. % in a timely fashion. 
The compiled circuit depth for the Aer simulation was about $300k$ {and the circuit depth for each each recurrent block (i.e. each time step) was about $27$ depending on the seed for the transpiler and weight initialization}. For statistical robustness, several seeds were tested and the best result from each bin was taken. Later, a noise model was compiled based off of the \texttt{ibm\_fez} device to serve as a test run for our IBM implementation.

Table \ref{tab:comparison_table} shows a comparison of using expectation values, the entire probability distribution, and the entire probability distribution under the \texttt{ibm\_fez} noise model simulator for various numbers of qubits. Figure \ref{fig:qrnn_test} depicts the full predictions of the $y$ and $z$ components overlayed on the true signal, along with the input signal and sampled reservoir features. It is shown that using the entire probability distribution of the output has better RMSE performance compared to the expectation value. Details of the simulation experiments are documents in the Supplementary Note \ref{SN:S3}. %While the performance is worse in the ``Distribution w. Noise'' test as expected, these results indicate that NISQ IBM hardware may also perform well, and motivates our next section.
%============ Figure 7 ==============
    \begin{figure*}[h!]
        \centering
        \includegraphics[width=\linewidth, keepaspectratio]{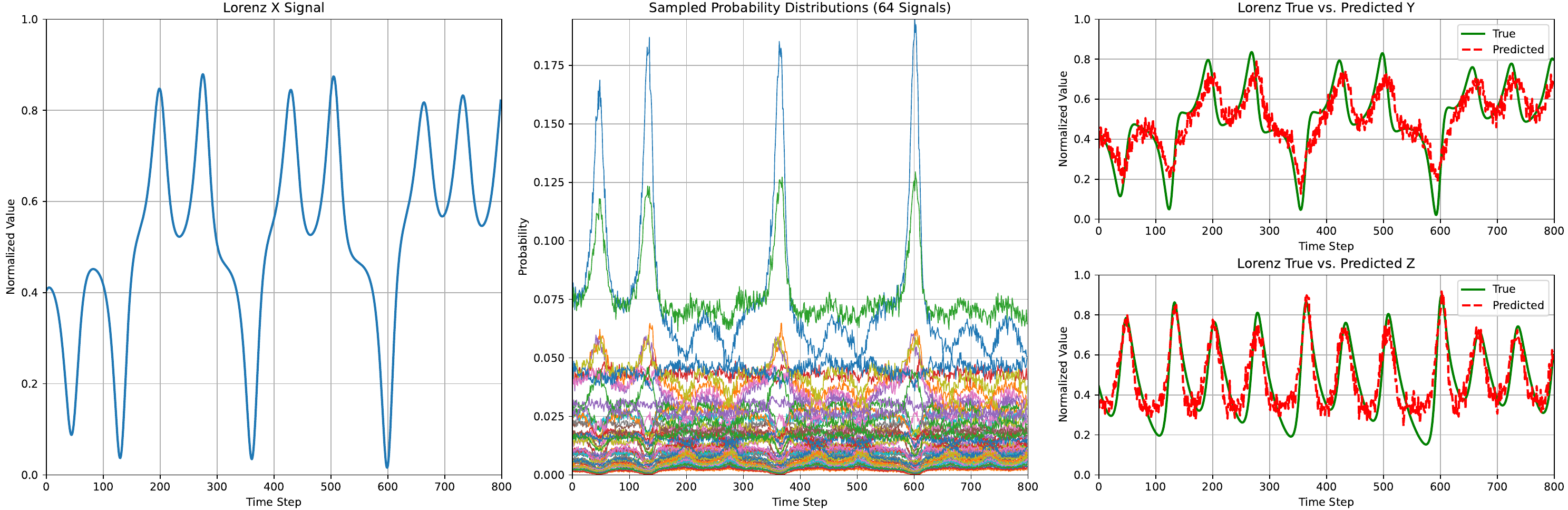}
        \caption{The full QPU test predictions of the $y(t+1)$ and $z(t+1)$ Lorenz system from the \texttt{ibm\_marrakesh} QPU with 12 qubits. From left-to-right the input signal, sampled reservoir features, and predictions are depicted, respectively. Input data is generated from the chaotic Lorenz system and only includes the $x(t)$ component. %For the hardware experiments, a smaller slice of the data was used to keep the IBM hardware from overloading and to minimize computational time. 
        Probability distributions of the QESN output states ${(2^6)}$, estimated by performing $60,000$ measurements of the circuit and aggregating the results over time are shown in the center panel. Shots are averaged over multiple runs, potentially introducing issues with ``drift'' of the quantum computer, but is necessary to prevent ``buffer overflow'' errors. At the end, the sampled probability distributions are mapped to the target signals using elastic net regression and the estimated and ground truth $y(t)$ and $z(t)$ trajectories are shown in the right panel.}
        \label{fig:qpu_test}
    \end{figure*}
%============ Figure 7 ==============

%=========== subsec: Comparison ==============================
\paragraph{Comparison with Classical Techniques:}\label{subsec: comparison with classical ESN}
%=========== subsec: Comparison ==============================
A brief experiment was conducted using a classical ESN that was trained and tested in direct comparison to the noiseless QESN. %to draw some conclusions about a possible quantum advantage. 
A basic linear regression model was also used in the comparison for sake of completeness. The test accuracy is shown to be lower on the QESN in all cases, as shown in Supplementary Note \ref{SN:S3} Figure \ref{fig:qrnn_comparison_aer}. This could be due to the inherent expressivity of qubits over bits, and the larger memory and feature space that they can provide when entangled. While there is always room for better hyperparameter tuning, our results indicate a consistent yet small performance gain from the noiseless QESN simulation in comparison to the classical ESN when the number of ESN reservoir nodes is fixed to be the same as the number of readout qubits.  We do not make any claims of quantum advantage at this point; however, we do note that future QESN designs or other architectures could greatly outperform the one presented in this paper and that the exponential growth of output features afforded by quantum computers could make this difference in performance even larger.

\section{Experimental Results}%{Hardware Implementation}
\label{subsubsec: hardware implementation}
To further test the efficacy of our QESN-based observer design, we conducted experiments on IBM superconducting qubit Quantum Processing Units (QPUs) and demonstrated state-of-the-art performance with gate-based QCs for time-series prediction. Our circuit was able to run coherently with a circuit depth of $\sim$200k, and demonstrated successful execution of over 1 million gates, all while producing robust features for well over the \texttt{ibm\_marrakesh} QPU's median $\mathcal{T}$1 and $\mathcal{T}$2 of 213.92 $\mu s$ and 119.57 $\mu s$ respectively. More specifically, this circuit was executed for no less than 48,000 $\mu s$, which is over 100 times the \texttt{ibm\_marrakesh} QPU's $\mathcal{T}$1 and $\mathcal{T}$2 times. This experiment proved that current NISQ computers are capable of %this sort of 
long time-series prediction in the present day. 

Our tests were performed on a time series of length 2000, greatly exceeding the previous length of 100 in \cite{yasuda2023quantumreservoircomputingrepeated} and 20 in \cite{Hu2024} by up to two orders of magnitude. These experiments were conducted using only 12 qubits, a number that was selected to create lower depth circuits with fewer SWAP operations, as well as reduce the potential of cross-talk errors. In contrast to the analog approach in \cite{kornjača2024largescalequantumreservoirlearning} and restart and rewind protocols indicated in \cite{Mujal2023}, this circuit ran recurrently with a coherent memory for its entire runtime, and did not require any stopping or resetting of the circuit, employing continuous mid-circuit measurements.

%The experimental setup for the hardware results differs slightly from the previously described simulated results. Due to some of the limitations of the IBM backend, the amount of data being tested was reduced from 9900 data points down to 2000. 
In addition, to overcome a previous known error on hardware implementations, which had also been identified in \cite{Hu2024} and described as the ``buffer overflow error'', we introduce a small delay to the circuit that creates a 24$\mu s$ delay immediately after each measurement. Doing this allowed us to mitigate this error, and enabled experimental run on much longer sequences of data. %Due to some opaque limitations in the IBM backend, reducing this delay further was not possible with the number of shots per job and number of data points we wanted to run. We do believe that this ``buffer overflow error'' is not something intrinsic to the QESN circuit design, but merits further investigation. It is likely that circuits of this depth with this number of measurements were not expected to be sent to the QPU in this fashion. 
% In order to fully avoid this problem, 
To further facilitate longer experimental runs, we split runs of the $2000$ data point circuit into jobs of $750$ shots, which was found to be the largest number of shots we could run per job with the given 24$\mu s$ delay and number of measurements. To aggregate back into a $60,000$ shot run like in the previous simulated experiments, we average the results over a total of $80$ jobs. 

Another important implementation difference between the simulation and experimental tests was in combining the train and test circuit into one. Instead of running each batch of data as a different circuit, a single circuit was run continuously for the 2000 data points, and then the averaged measurements were split into train and test sets of 1200 and 800 after the fact, respectively. A washout length of only 15 was removed from the train data in the hardware experiments, as we observed in the measurement data that this would be sufficient. Typically, washout length is a symptom of fading memory and the echo-state property, and we observe this washout in the output signals recovered from the QPU. In Supplementary Note \ref{SN:S3} Figure \ref{fig:qpu_exp_values}, clear gradual initialization is shown for the first 15 data points, indicating the presence of a fading memory on hardware.

Figure \ref{fig:qpu_test} shows the full prediction from the \texttt{ibm\_marrakesh} QPU alongside the input signal and recorded reservoir signals. Getting a coherent output over a data length of 2000 with a circuit depth of 200k has not been demonstrated on NISQ hardware given the short $\mathcal{T}$1 and $\mathcal{T}$2 of a standard qubit. The signals recovered from the QPU demonstrated relatively high fidelity for NISQ hardware and were used fit to the target signals $y(t+1)$ and $z(t+1)$ using a regularized optimization process. This process gave allowed for a robust hardware demonstration of long time-series prediction on IBM QPU's using just 12 qubits. The RMSE was .0707 for train and .0922 for test respectively, falling closely inline with the expected noisy simulated results.

The sampled Lorenz system proved a particularly difficult task for the NISQ hardware due to the fine granularity of the time discretization used in our dataset, which required highly-precise output features from time-step to time-step. Even with the difficulties of current NISQ hardware, our circuit shows clear predictive capabilities beyond the training data, and a strong signal in the presence of noise, whether it be from quantum gate-errors, or other channels. We do note that while the circuit did run for 48,000 $\mu s$ per shot, the deterministic delay included for overcoming the implementation errors contributed nontrivially to this run time. While this delay enabled us to collect more measurements than would typically be possible, it also artificially inflated the circuit's runtime, reducing the overall efficiency of the process. Despite this, the delay facilitated the experiment, allowing us to observe long-term coherence with a circuit that ran for over $100$ times the $\mathcal{T}_1$ and $\mathcal{T}_2$ times of the \texttt{ibm\_marrakesh} QPU. 
%While this delay did allow us to get more measurements than would typically be possible, it also artificially inflates the runtime of the circuit, and makes the entire process less efficient. That being said, it is the delay itself that allowed us to conduct this experiment, and it is also the reason why we can say definitively that we have demonstrated long-term coherence with a circuit that ran for over 100 times the $\mathcal{T}$1 and $\mathcal{T}$2 of the \texttt{ibm\_marrakesh} QPU.

%============ Figure 9 ==============
    % \begin{figure}[h!]
    %     \centering
    %     \includegraphics[width=\linewidth, keepaspectratio]{figures/QPU-probs-values.pdf}
    %     \caption{Probability distribution over time recorded from the \texttt{ibm\_marrakesh} QPU. Washout of length 15 has been removed for visibility.}
    %     \label{fig:qpu_probs_values}
    % \end{figure}
%============ Figure 9 ==============

%=============== sec: Conclusions ==================
\section{Discussions}\label{sec:conclusions}
%=============== sec: Conclusions ==================
In this work, we introduced quantum observers based on QESNs for predicting the latent variables of dynamical systems using partially observed system variables. The QESNs, inspired by classical echo-state networks, incorporated sparser gate weights, which not only reduced computational complexity but also minimized gate operations, proving to be a valuable design parameter for constructing quantum circuits optimized for hardware. To systematically design sparser weights and analyze the QESN circuit, we extended classical time-domain response analysis principles to the quantum domain, applying them to the QESN architecture to characterize its dynamic behavior. This analysis provided deeper insights into the temporal evolution of quantum states in response to structured inputs. The rise time, settling time, and condition number of the feature signals generated by the QESNs offer a systematic approach for assessing the fading memory, coherence, and nonlinear characteristics of the quantum circuit. Furthermore, in addition to the sparsity hyperparameter, data re-uploading technique was introduced, demonstrating their impact on controlling circuit depth, gate count, and nonlinearity. 

The resulting quantum observer performed well in both simulation and hardware experiments on the prediction of the chaotic Lorenz system.  We demonstrate state-of-the-art performance on IBM hardware and run the circuit continuously with persistent memory and coherent outputs for over 100 times the \texttt{ibm\_marrakesh} QPU's median $\mathcal{T}$1 and $\mathcal{T}$2 of 213.92 $\mu s$ and 119.57 $\mu s$ to demonstrate, to the best knowledge of the authors, the longest-ever run time-series prediction algorithm on gate-based NISQ hardware.

These results underscore the potential of quantum observers based on QESNs as a powerful tool for dynamic system prediction, even on current NISQ hardware. The ability to run time-series prediction algorithms for extended durations on quantum processors demonstrates the practical viability of quantum machine learning approaches today. With further refinements in quantum hardware, including fault-tolerant qubits and improved connectivity, we anticipate that quantum observers like the QESN will scale to address increasingly complex problems, including those in high-dimensional and nonlinear systems. This work sets a promising foundation for future research into quantum-enhanced machine learning and opens up new avenues for solving real-world dynamical challenges that were previously thought to be beyond the reach of today's quantum technologies.

\section{Acknowledgments}
The authors EC and VN would like to acknowledge the funding support from NSWC, Indian Head, under award \# N00174-23-1-0006. ENE would like to acknowledge funding from the Naval Innovation Science and Engineering (NISE) program. The entire team would like to acknowledge the support of the IBM Quantum team for providing valuable feedback and suggestions in the process of making these experiments run on hardware. Any opinions, findings,
conclusions, or recommendations expressed here are those of the authors and do not necessarily reflect the views of the aforementioned groups.


\section{Data Availability:}
The datasets generated and/or analyzed during the current study are available in the QESN-Code repository, \href{https://github.com/econnerty/QESN-Code/}{https://github.com/econnerty/QESN-Code/}.
\subsection{Code Availability:}
The underlying code and training/validation datasets for this study is available in the QESN-code repository, and can be accessed via this link \href{https://github.com/econnerty/QESN-Code/}{https://github.com/econnerty/QESN-Code/}. 
%\printbibliography
\bibliography{sn-bibliography}% common bib file
%\input main-nature.bbl
% \newpage
\appendix
% \begin{appendices}
\section{Supplementary Note 1}\label{SN:algorithm}
%=================== Sec: Prelims ============================
\subsection{System Architecture}\label{sec:notations}
%=================== Sec: Prelims ============================
\subsubsection{Preliminaries}
The input data is denoted by 
% $X = \{x_t\}_{t=1}^N$
$X= \{x_1, \dots, x_t, \dots, x_N\}$
, with $x_t \in \mathbb{R}^d$ and $t \in T = \{1, \dots, N\}$. We encode the classical data onto $n_q$ total qubits, which must be even, using a context window $X_c = \{x_{t-c}, \dots, x_t\}$ of length $c$. We utilize a re-uploading scheme to introduce nonlinearity with $n_c$ repeated circuit blocks, and treat $n_c$ as a tunable hyperparameter that can increase the amount of nonlinearity in the circuit. %The qubits are grouped into two equal registers, 
The $\textit{Memory}$ and $\textit{Readout}$ registers consist of $\frac{n_q}{2}$ qubits each. The set of pairs of memory and readout qubits are denoted as $P = \{P_i\}_{i=1}^{n_q/2}$ with $P_i = (R_i, M_i)$ denoting the $i$th element in the pair consisting of readout $R_i$ and memory $M_i$. We similarly denote the set of memory qubits $M$ and the set of readout qubits $R$.

We denote the random weight matrices $\mathbf{W}_\text{in} \in \mathbb{R}^{cd \times n_q  \times 3}$,
%{d \times n_q \times c \times 3}$
$\mathbf{W}_\text{bias} \in \mathbb{R}^{n_q \times 1}$, $\mathbf{W}_\text{ent} \in \mathbb{R}^{\frac{n_q}{2} \times 2}$, and $\mathbf{W}_\text{mem} \in \mathbb{R}^{\frac{n_q}{2} \times 1}$, which are used for rotations and their biases, as well as for entangling gates between qubits, respectively. Denote the single-qubit rotation gate $\mathbf{R(q, \alpha, \beta, \gamma)}$ applied to $\mathbf{q}$, with the rotation angles $\alpha,\beta,\gamma \in \mathbb{R}$.
% are computed from the context window and the weight matrices $\mathbf{W}_\text{in}$ and $\mathbf{W}_\text{bias}$. 
The gates used in the QESN (shown as part of Figure \ref{fig:QRC_Circuit}) include Controlled-X (C-NOT), Controlled-RY (CRY), Controlled-RX (CRX), and Controlled-RZ (CRZ). Finally we denote sparsity parameter $\kappa \in [0,1]$. 

%Finally, the readout qubits \cred{are measured by an observable operator denoted $\mathcal{O}$}.

%=================== subsec: Sys architecture ========================
\subsubsection{Quantum Echo-state Network}\label{subsec:system-architecture}
%=================== subsec: Sys architecture ========================

Here we give the %a general overview of the QESN circuit pictured in Figure \ref{fig:QRC_Circuit}, as well as 
pseudocode of the proposed QESN in Algortihm \ref{qesnalgo}. %and the overall pipeline from start to finish of both the classical and quantum components of the algorithm in Figure \ref{fig:QRC_Pipeline}.

%============== Algorithm ===================
\begin{algorithm}
\caption{QESN Algorithm}
\small
\begin{algorithmic}
% \begin{algopseudocode} # environment is undefined
    \Procedure{Create\_Quantum\_Circuit}{$\mathbf{X}, n_q, c, n_c, \kappa$}
    \State \textbf{Initialize:} 
    %\State Set seed $s$. 
    %\State Verify $n_q$ is even
    \State Create \textit{Memory} and \textit{Readout} registers with $\frac{n_q}{2}$.
    \State Initialize all qubits to $\ket{0}$
    \State Initialize random weights $\mathbf{W}_\text{in}$, $\mathbf{W}_\text{ent}$, and $\mathbf{W}_\text{mem}$.
    \State Apply sparsity to $\mathbf{W}_\text{ent}$ and $\mathbf{W}_\text{mem}$ using $\kappa$
    \For{$t = c$ \textbf{to} $N-c$}
        \State Extract context window $\mathbf{X_c} \gets \mathbf{X}[t-c:t]$
        \For{$j = 1$ \textbf{to} $n_c$}
            \State \textbf{Parallel operations on qubit pairs}:
        %\ForEach{$(q_i, q_j)$ in paired qubits}
            % \State $\alpha_1, \beta_1, \gamma_1 \gets W_{in}^{P_1} \cdot \mathbf{X_c} + W_{bias}^{P_1}$
            % \State $\alpha_2, \beta_2, \gamma_2 \gets W_{in}^{P_2} \cdot \mathbf{X_c} + W_{bias}^{P_2}$
            \State $\alpha_i, \beta_i, \gamma_i \gets W_{in}^{P_i} \cdot \mathbf{X_c} + W_{bias}^{P_i},\quad$ $i=1,2$
            \State $\epsilon_1,\epsilon_2 \gets W_{ent}^P$
            \State Apply rotation: $R({P_i}, \alpha_i, \beta_i, \gamma_i)$, $\quad i=1,2$  % $R({P_2}, \alpha_2, \beta_2, \gamma_2)$
            % \State Apply rotation: $R(\cred{P_2}, \alpha_2, \beta_2, \gamma_2)$
            \State Apply C-NOT gate: C-NOT$(P_1, P_2)$
            \State Apply rotation: $R(P_i, \alpha_i, \beta_i, \gamma_i)$, $\quad i=1,2$ %$R(P_2, \alpha_2, \beta_2, \gamma_2)$
            % \State Apply rotation: $R(P_2, \alpha_2, \beta_2, \gamma_2)$
            \State Apply controlled-RY gate: CRY$(P_1, P_2, \epsilon_1)$
            \State Apply rotation: $R(P_i, \alpha_i, \beta_i, \gamma_i)$, $\quad i=1,2$ %$R(P_2, \alpha_2, \beta_2, \gamma_2)$
            % \State Apply rotation: $R(P_2, \alpha_2, \beta_2, \gamma_2)$
            \State Apply controlled-RX gate: CRX$(P_1, P_2, \epsilon_2)$
            \State \textbf{Sequential operations on memory qubits}:
            \State $\epsilon_3 \gets W_{mem}^M$
            \State Entangle memory qubits: CRZ($M_1$, $M_2$, $\epsilon_3$)
        \EndFor
        \State Measure and reset qubits in \textit{Readout} register to $\ket{0}$.
    \EndFor
    \State \textbf{Return:} QESN Circuit
\EndProcedure
% \end{algopseudocode} # not recognized
\end{algorithmic}
\label{qesnalgo}
\end{algorithm}
%============== Algorithm ===================
%Weight initialization plays an important role in the dynamics of the classical ESN. 
Randomly initialized weights play an important role in creating reservoir-like behavior in the QESN. This approach mimics the inherent randomness and sparsity of ESNs, but with quantum computers. The input weights are sampled from a normal distribution $\mathcal{N}(\mu, \Sigma)$, truncated to the interval $(0, \pi]$. Input weights are then scaled by the size of the context window and the number of repeated blocks within the QESN to ensure that the rotation from any single data point does not greatly perturb the system. Sparsity is injected into the two qubit gate connections, excluding the C-NOT gate, which is always present, making some percentage of the entangling gate weights zero, mimicking the sparse connectivity found in classical ESNs, and can reduce the number of potential gate errors as well as the depth of the circuit. 

\section{Supplementary Note 2}\label{SN:response-analysis}
%================== sec: response analysis ===================
\subsection{Response Analysis}\label{subsec:response_analysis}
%================== sec: response analysis ===================
In this section, we analyze the response of the QESN circuit to three basic signal types: sinusoid, step, and ramp. In order to gain a better understanding of the circuit's dynamics, these basic signals were input into the QESN circuit and measured as both expectation values and sampled probabilities over the computational basis states. Namely, these tests aimed to demonstrate the empirical presence of \textit{memory} and \textit{nonlinearity} in the circuit under different sparsity levels, as well test the effect of our ``data re-uploading block'' hyperparameter. Sparsity, a tunable hyperparameter introduced in our QESN circuit allows for reduced circuit depth and gate complexity, as well as reduced number of SWAPs on IBM hardware when it is applied to two-qubit gates on non-adjacent qubits. In addition, ``data re-uploading blocks'' give tunable amounts of nonlinearity in output features. Together, these provide desirable features for our QESN that may improve performance, especially on IBM hardware. We show empirically that some degree of sparsity is not only good for circuit depth, but also appears to improve various metrics such as the condition numbers of the output feature matrices. We also show the effect of the repeatable ``data re-uploading block" on the set of input signals, demonstrating the difference each repeated block makes on the output.

For the sparsity tests, different degrees of sparsity, expressed as percentages, were tested using the same set of randomized weights. The percentages were calculated by counting up all of the entangling gates in the randomly generated circuits and dividing by the total number of possible entangling gates in Figure \ref{fig:QRC_Circuit}, including C-NOT, and were as follows: (a) 0\% sparsity, or the exact circuit in Figure\ref{fig:QRC_Circuit} and Algorithm   \ref{qesnalgo}, extrapolated to 12 qubits, (b) 29\% sparsity, a circuit that had some gates randomly zeroed out, (c) 42\% sparsity, a circuit similar to the previous one except no CRZ gates were present, indicating no interactions between memory qubits, and (d) 100\% sparsity, a circuit without any entangling operations, and therefore no interaction at all between memory and readout. Using these tests we demonstrate the presence of \textit{memory} and \textit{nonlinearity} in our circuit, as well as the modularity of our design and robustness to randomized entangling schemes.

%----------------------%
\textbf{Memory:}
%---------------------%
Memory is a requirement for any RNN \cite{learninginternal} or ESN \cite{jaeger2001} type neural network, thus it is important for us to demonstrate this property in QRNNs, which must also have a way to store a hidden representation of past inputs. An easy way to demonstrate this property is to input a simple ``step" signal into the QESN circuit and identify the rise-time of the measured outputs w.r.t. to the input signal. For a more detailed mathematical formalism involving weak or quantum nondemolition (QND) measurements and \textit{fading memory} through volterra series analysis, we refer to \cite{Hu2024, yasuda2023quantumreservoircomputingrepeated, chen2020temporal}. In addition to leaning on this formalism, we show strong evidence via numerical simulations that our QESN circuit does indeed exhibit a \textit{fading memory} through different sparsity levels as long as there some entanglement in the circuit. The only scenario in which the circuit has no entanglement is when sparsity is taken to its extreme value of 100\%.

\begin{figure*}[h]
    \centering
    \includegraphics[width=\linewidth, keepaspectratio]{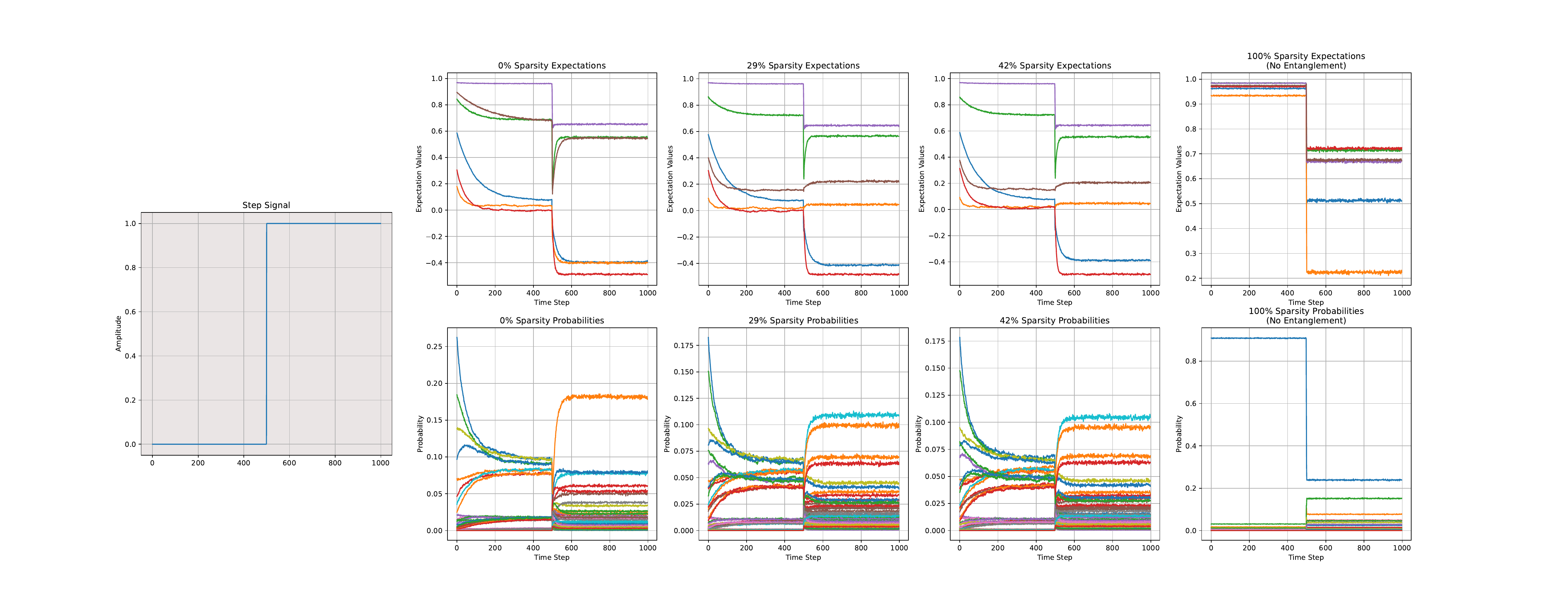}
    %\caption{Step signal processing by the QESN circuit, with a focus on the rise-time and memory introduced by varying the entanglement and sparsity configurations.}
    \caption{Step signal processing by the QESN circuit, with a focus on the rise-time and memory introduced by varying the sparsity configurations. Top row of features are Pauli-Z expectation values, and the bottom row are full $2^n$ feature signals derived from the probability distribution of each computational basis state. Of note is the non-zero rise time in the signal during transition from initial state to the fully saturated ``0'' input state, and the subsequent transition when the ``1'' input is applied in the step signal. This behavior is present in all configurations except for the 100\% sparsity case, which has an instant response time due to the lack of entanglement, and therefore a lack of memory.}
    \label{fig:step_response}
\end{figure*}

Figure \ref{fig:step_response} shows a non-zero rise-time in the QESN expectation values and probability distribution outcomes in the presence of a sudden increase in signal input from zero to one. We also note that the apparent change in dynamics in the first half of the signal is due to the QESN's bias weights, which introduce small rotations on each qubit at each timestep even when the input is zero. From this, we see a non-zero rise-time from the QESN's initial state to its first equilibrium state, typically called the ``washout length'' in classical reservoir networks \cite{jaeger2001,miao2022interpretable}, and then another non-zero rise-time to a new equilibrium state under the presence of the constant unit input of the step signal halfway through. 

These non-zero rise-times to new states are explained by the presence of a persistent memory within the QESN circuit. Interestingly enough, this behavior changes drastically in the ``100\%'' sparsity case. When the memory and readout qubits in the QESN are completely decoupled, we see behavior that is much more akin to a classical feed-forward neural networks rather than RNNs. As long as there is some degree of entanglement within our circuit between memory and readout qubits, we have shown there exists a fading memory in our circuit that is tolerant to random weight initialization and different sparsity schemes.

%-----------------------%
\textbf{Nonlinearity:}
%-----------------------%
Continuing on with these simple signals, we aim to demonstrate the existence of \textit{nonlinearity} in our input-to-output mapping. It is trivial to show that real valued inputs embedded onto a quantum system as a sequence of rotations are nonlinear due to the implied sinusoidal activation of angle embeddings \cite{Govia_2022}, but it is still important to analyze this effect in the presence of more advanced or complex entanglement schemes. Namely, we would like to demonstrate the ``richness" of quantum reservoir circuits in both their expectation value output and their full $2^n$ sampled probability distribution output. We do this by inputting a simple ramp and sinusoid signal into the QESN as done before with the step signal, and instead of looking for rise time, we examine the higher order harmonics of the resulting feature signals as shown in the sinusoid, and the very clear nonlinear trajectories demonstrated in the ramp signal. We also examine the condition numbers of the output feature matrices in each case.

It should be noted that in the context of our regression problem, the condition number of the feature matrix is a measure of the ratio of largest to smallest singular values of the feature matrix, which is connected to the matrix's invertability. Thus lower values indicate a decrease in sensitivity of the QESN to small changes in input and are correlated with feature richness. We measure condition numbers for the features derived from the sampled probability distribution outputs to show clear improvement in numerical stability with our entanglement scheme, which is not significantly affected by sparsity. The condition numbers for each sparsity percentage and each signal are shown in Table \ref{tab:condition_number}. As long as entanglement operations are present, a relatively low condition number can be obtained.

\begin{figure*}[h]
    \centering
    \includegraphics[width=\linewidth, keepaspectratio]{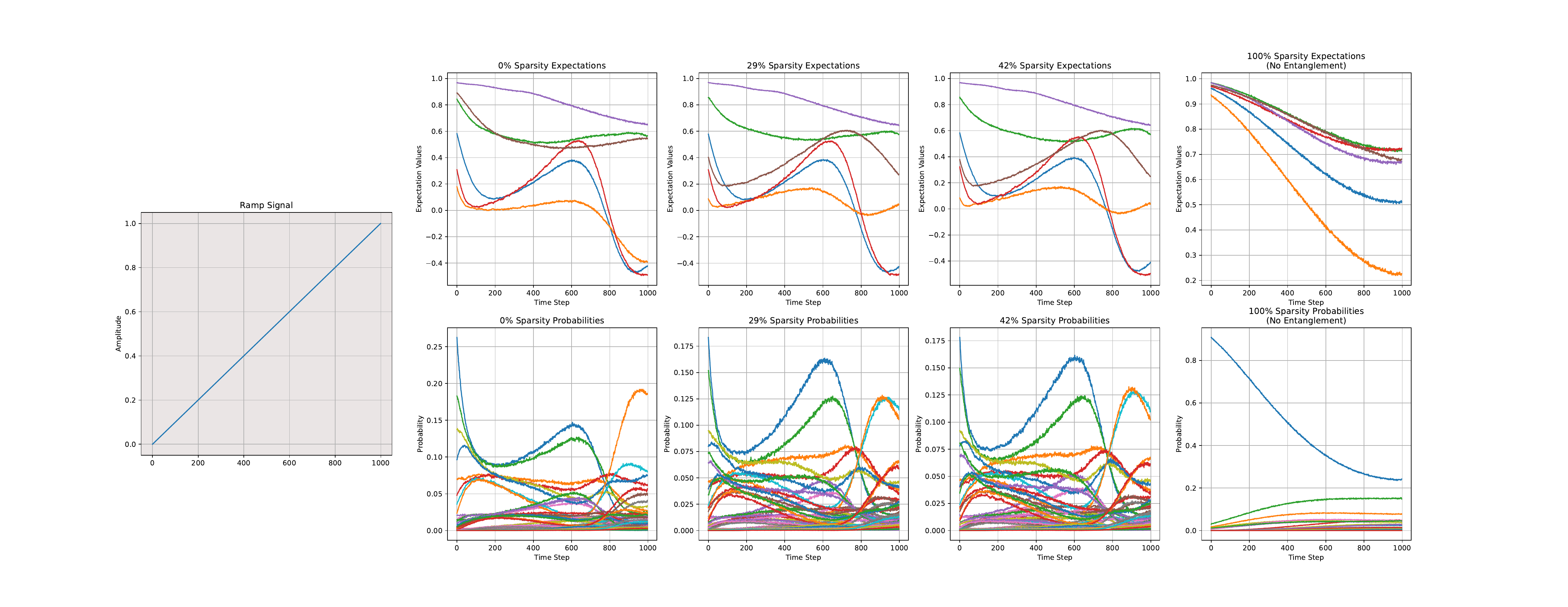}
    \caption{Analysis of a ramp input signal processed by the QESN circuit, illustrating the dynamic response and nonlinear mappings as a function of sparsity and entanglement. Top row of features are Pauli-Z expectation values, and the bottom row are full $2^n$ feature signals derived from the probability distribution of each computational basis state.}
    \label{fig:ramp_response}
\end{figure*}

\begin{figure*}[h]
    \centering
    \includegraphics[width=\linewidth, keepaspectratio]{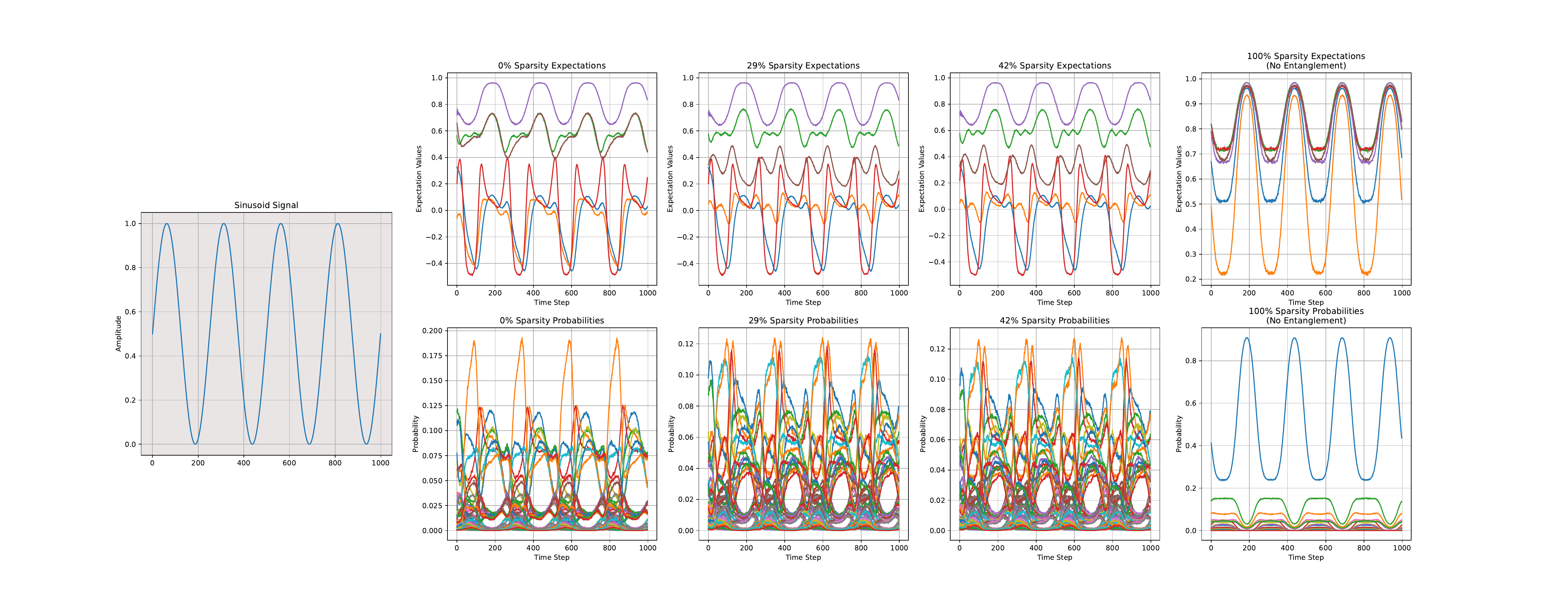}
    \caption{Response of the QESN circuit to a sinusoid input signal across different configurations, highlighting the richness of features with varying levels of sparsity and entanglement. Top row of features are Pauli-Z expectation values, and the bottom row are full $2^n$ feature signals derived from the probability distribution of each computational basis state.}
    \label{fig:sinusoid_response}
\end{figure*}

Figure \ref{fig:ramp_response} shows a ramp signal input into the QESN over different sparsity levels and demonstrate rich nonlinearity extracted from a simple 1-D input. Figure \ref{fig:sinusoid_response} shows a similar test done using a sinusoidal signal, with higher order harmonics present in all cases where entanglement is present, showing sparsity is not detrimental to performance. It is clear from these plots that 1.) rich nonlinearities are present in the input-to-output mapping, and 2.) sparsity is a hyperparameter that can be tuned to relatively high values without sacrificing the circuit's expressivity. Demonstrating these capabilities allow us to find optimal parameters for both our simulated and IBM hardware implementation of the QESN circuit.

%-----------------------%
\textbf{Re-uploading Blocks:}
%-----------------------%
Repeatable ``data re-uploading blocks" are a tunable hyperparameter in our QESN circuit that take inspiration from \cite{P_rez_Salinas_2020,chu2022qmlperrortolerantnonlinearquantum} which describe the ability of QNN performance to increase with a ``data re-uploading" technique that places repeated input guided rotations between two-qubit entangling gates. Taking this idea, we allow our inner circuit block from Figure \ref{fig:QRC_Circuit} to be repeated an arbitrary number of times as specified by the user. While this will increase circuit depth, we have found the trade-off to be worth it.

\begin{figure*}[h]
    \centering
    \includegraphics[width=\linewidth, keepaspectratio]{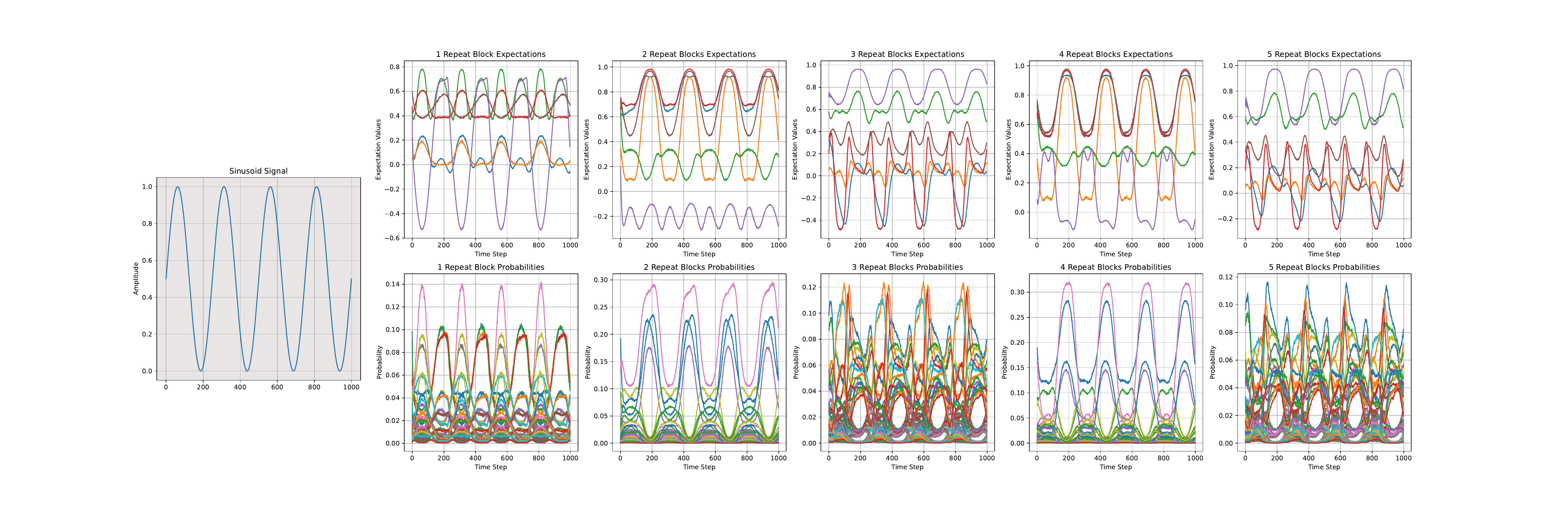}
    \caption{Response of the QESN circuit to a sinusoid input signal across differing number of repeat blocks. Top row of features are Pauli-Z expectation values, and the bottom row are full $2^n$ feature signals derived from the probability distribution of each computational basis state. In our experiments, $n_c = 3$ was ultimately chosen as the number of repeat blocks due to the higher amounts of nonlinearity in the output. We do note that the seeming decrease in performance seen with even number of repeat blocks is likely due to the C-NOT gate in our circuit ``dominating" the block and inverting itself on even counts. In these cases, it seems both the memory and expressivity are hampered.}
    \label{fig:sinusoid_response_rep_blocks}
\end{figure*}

\begin{figure*}[h]
    \centering
    \includegraphics[width=\linewidth, keepaspectratio]{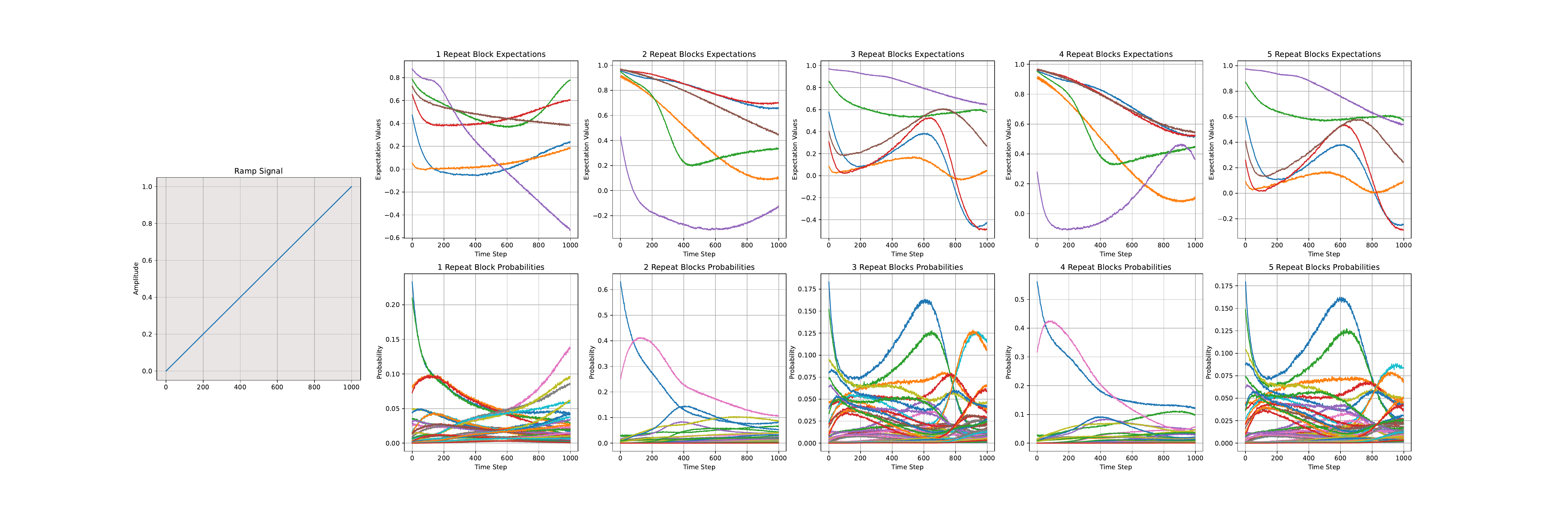}
    \caption{Response of the QESN circuit to a ramp input signal across differing number of repeat blocks. Top row of features are Pauli-Z expectation values, and the bottom row are full $2^n$ feature signals derived from the probability distribution of each computational basis state.}
    \label{fig:ramp_response_rep_blocks}
\end{figure*}

\begin{figure*}[h]
    \centering
    \includegraphics[width=\linewidth, keepaspectratio]{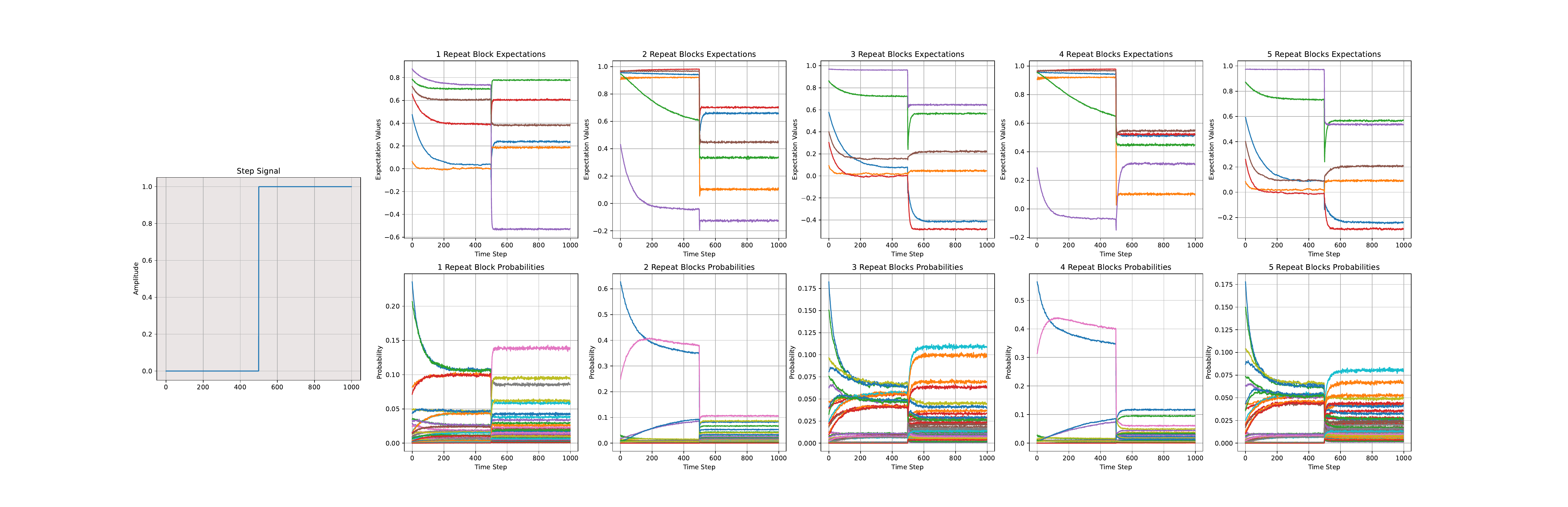}
    \caption{Response of the QESN circuit to a step input signal across differing number of repeat blocks. Top row of features are Pauli-Z expectation values, and the bottom row are full $2^n$ feature signals derived from the probability distribution of each computational basis state.}
    \label{fig:step_response_rep_blocks}
\end{figure*}

The sinusoidal signal is shown as a function of number of repeat blocks in Figure \ref{fig:sinusoid_response_rep_blocks}, and shows that increasing odd numbered repeat blocks seem to increase the amount of nonlinearity in our input-to-output mapping. The seeming decrease of performance on even number of repeat blocks is thought to arise from the dominance of the C-NOT gate in our QESN circuit shown in Figure \ref{fig:QRC_Circuit}, which is the strongest entangling operation in the circuit and may be inverting itself on even numbers of runs. This is likely due to the relatively low magnitude of weights associated with the other two-qubit gates in the circuit, as well as the sparsity. This behavior is also present in the even repeats of Figure \ref{fig:ramp_response_rep_blocks} and Figure \ref{fig:step_response_rep_blocks}, indicating both a reduction in memory capacity and expressivity. It should be noted that this may be due to the particular weight initializations that we have chosen and may not hold true in all circuit's following this gate sequence. With this, we conclude that the C-NOT gate is critical to our particular design, contributing greatly to the QESN's \textit{memory}, \textit{nonlinearity}, and overall expressivity. We ultimately choose $n_c = 3$ as the number of repeat blocks used for all of our experiments on the Lorenz system, as it appears to yield higher feature richness and does not dramatically increase the depth of the circuit.

%We also show similar behavior in the ramp and step signals as seen in Figures \ref{fig:step_response_rep_blocks} and \ref{fig:ramp_response_rep_blocks}, demonstrating consistent behavior across different inputs with the QESN.

These experiments validate our hypothesis that the QESN circuit can effectively extract and amplify nonlinear characteristics from time-series data, an essential feature for advanced predictive modeling in complex dynamical systems. We demonstrate this capability even in the presence of sparsity, as shown by the similarity of outputs in the sparse versions of the circuit. This is also shown in Table \ref{tab:condition_number}, which shows better condition numbers with sparse circuits versus the alternative. Overall, lower condition numbers and visual validation of the output feature robustness provide enough evidence that sparsity can be exploited as a way to reduce circuit depth and decrease the probability of errors on NISQ hardware. In addition, the tunable hyperparameter for repeatable ``data re-uploading blocks'' can further increase the degree of nonlinearity in our input-to-output mapping, and therefore allows for more complex relationships to be learned.

\begin{table*}[h]
    \centering
    \begin{tabular}{l c c c c}
        \hline
        \textbf{Signal Type} & \textbf{0\% Sparsity} & \textbf{29\% Sparsity} & \textbf{42\% Sparsity} & \textbf{100\% Sparsity} \\
         &  &  &  & \textbf{(No Entanglement)} \\
        \hline
        Ramp & 2997.84 & 2581.41 & \textbf{2529.65} & 40729.79 \\
        Sinusoid & 2977.74 & \textbf{2235.64} & 2259.78 & 38777.81 \\
        Step & 4470.22 & \textbf{2696.22} & 2707.04 & 35337.63 \\
        \hline
    \end{tabular}
    \caption{Condition Numbers Across Different Sparsity Levels. Lower is better.}
    \label{tab:condition_number}
\end{table*}

\section{Supplementary Note 3}
\label{SN:S3}
%==========================

%============= subsec: lorenz_system=============
% \subsection{Lorenz System}\label{subsec:lorenz_system}
%==============subsec:lorenz_system==============

%================== subsec: Data handling and training =================
\subsection{Data Handling and Training}\label{subsec:data-training}
%================== subsec: Data handling and training =================
Training data is created from numerical simulations of the Lorenz system (shown in Figure \ref{fig:QRC_Pipeline}), a well-known chaotic system often used to test the performance of predictive models. The Lorenz system is defined by the following set of differential equations:
\begin{align*}\label{eq:lorenz}
    \frac{\rm{d}x}{\rm{d}t} = \sigma (y - x), \quad
    \frac{\rm{d}y}{\rm{d}t} = x (\rho - z) - y, \quad
    \frac{\rm{d}z}{\rm{d}t} = xy - \beta z \text{,}
\end{align*}
where $\sigma = 10$, $\beta = \frac{8}{3}$, and $\rho = 28$ are parameters that define the behavior of the system. To ensure the consistency of the QESN's responsiveness, the data is normalized to values $\in [0,1]$. The training set consists of 9900 data points, split into training and test sets of 6900 and 3000 data points, respectively, where each data point contains the $x(t)$, $y(t)$, and $z(t)$ variables of the Lorenz system. In our experiments, only the $x(t)$ was fed into the QESN, with the task being to predict $y(t+1)$ and $z(t+1)$ given this single signal. This setup tests the QESN's ability to learn and predict data generated by chaotic, nonlinear dynamics. %of a chaotic system.

{First, the QESN circuit is run on the training data, which produces a set of output signals from quantum measurements. This output signal is then used to fit a regression model. The first 300 data points of the output signal are discarded as is typically done in classical ESNs. This discarded data is typically known as the washout, which is the length of time needed to forget initial conditions. Finally, the QESN circuit is run on the test data, and the learned weights are used to perform predictions. Elastic net regularization \cite{elastic_net} is used to prevent overfitting for each test case, which introduces regression hyperparameters that are tuned for each case to minimize test error.}

\subsection{Supplementary Results}\label{subsec:supplementary_results}
\begin{figure}[h]
    \centering
    \includegraphics[width=\linewidth, keepaspectratio,trim={0cm 0 0 1.4cm},clip]{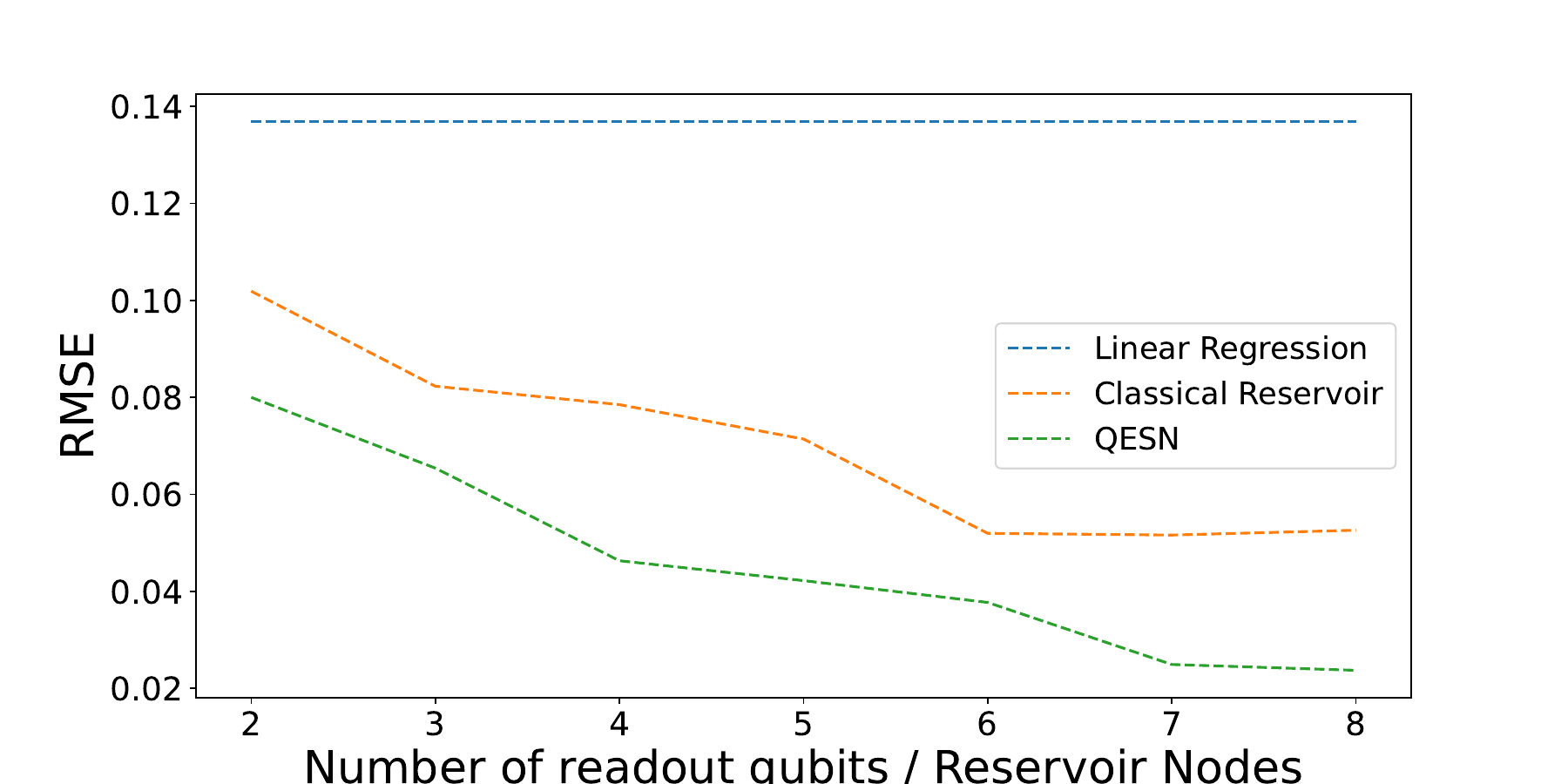}
    \caption{Test set loss with classical reservoir, basic linear regression, and QESN architecture.}
    \label{fig:qrnn_comparison_aer}
    \vspace{-0.3cm}
\end{figure}

In this section, we detail more of the results gathered from numerical simulations and hardware experiments. As detailed above in Section \ref{subsec: comparison with classical ESN}, Figure \ref{fig:qrnn_comparison_aer} shows a brief comparison of the QESN architecture with a classical reservoir network. We highlight that the QESN is already showing signs of an advantage when the number of output nodes is set to be the same. As fault-tolerant qubits become available, we expect this to allow for even more in depth experiments. 

\begin{figure}[h]
        \centering
        \includegraphics[width=\linewidth, keepaspectratio]{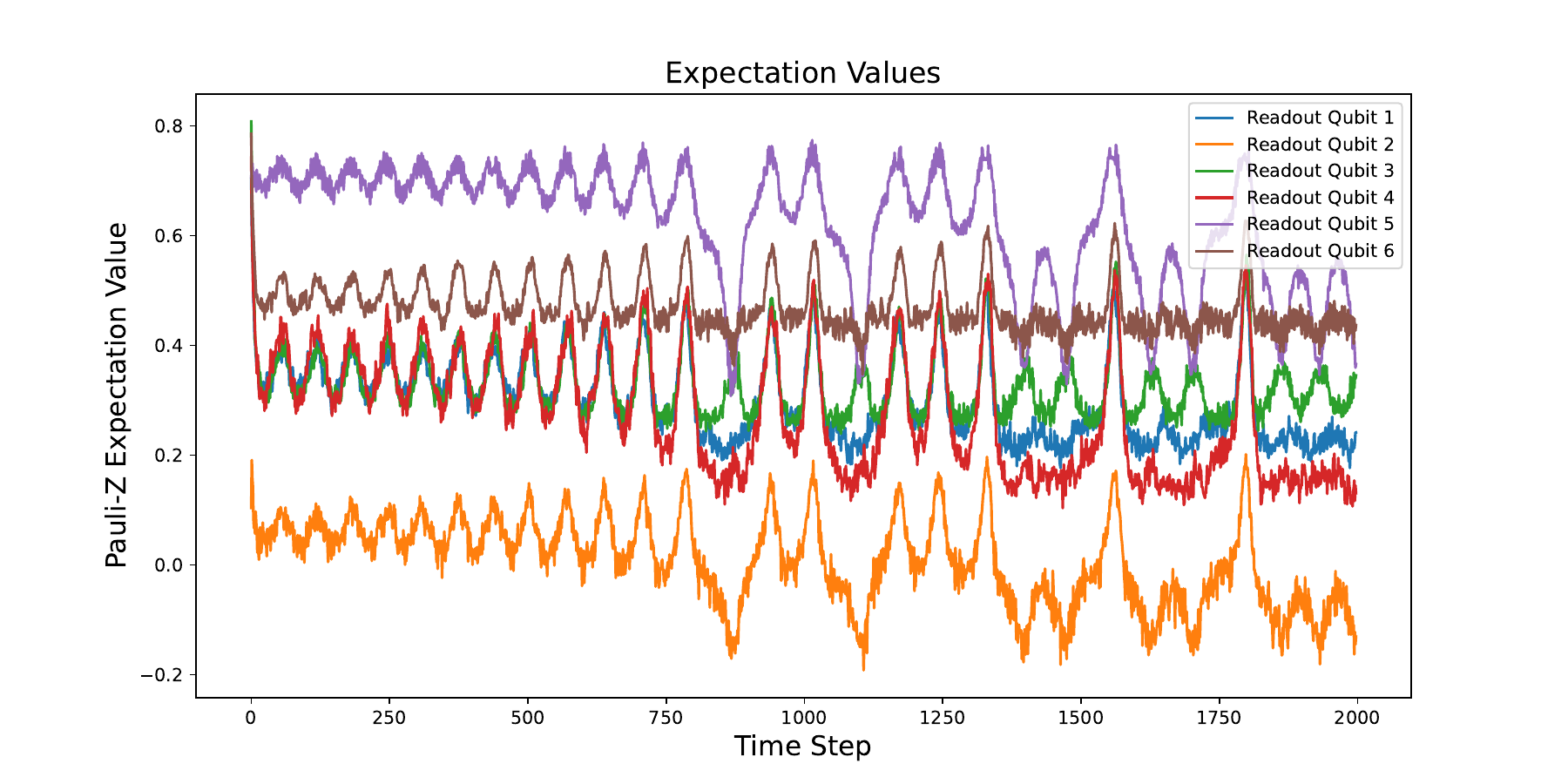}
        \caption{Expectation values recorded from the \texttt{ibm\_marrakesh} QPU. A washout length is present in the initial 15 data points, indicating QESN memory on the IBM hardware.}
        \label{fig:qpu_exp_values}
    \end{figure}

In addition, we give an example of clear memory capacity in Figure \ref{fig:qpu_exp_values} on IBM Quantum hardware. We make the observation that in the Pauli-Z expectation value of the readout qubits tested on the \texttt{ibm\_marrakesh} QPU, there is a clear gradual initialization of the reservoir that last for roughly 15 time steps. As demonstrated in our response analysis in the noiseless environment, this result is in strong agreeance with our numerical results and demonstrates clear memory capacity within the quantum circuit on actual hardware. As noted before, non-zero response times to new signals indicate that our QESN has a memory capacity, which is a product of the partial \textit{measure and reset} paradigm.

%============ Figure 10 ==============
%============ Figure 10 ==============

%============ Figure 7 ==============
%============ Figure 7 ==============

\end{document}